\documentclass[nohyper,12pt,letterpaper]{JHEP3}
\usepackage{epsfig,youngtab}

\author{Robert de Mello Koch$^{1,2}$, Jelena Smolic$^{1}$ and Milena Smolic$^{1}$\\
\qquad \\
$^{1}$Department of Physics and Centre for Theoretical Physics,\\ 
University of the Witwatersrand,\\ 
Wits, 2050,\\ 
South Africa\\
\qquad\\
$^{2}$Stellenbosch Institute for Advanced Studies,\\
Stellenbosch,\\
South Africa\\
\qquad\\
E-mail: \email{robert@neo.phys.wits.ac.za, smolicj@science.pg.wits.ac.za, smolicm@science.pg.wits.ac.za}}

\abstract{
We study the one-loop anomalous dimensions of operators in the ${\cal N}=4$ super Yang-Mills theory
that are dual to open strings ending on giant gravitons. We consider both AdS and sphere giants as 
well as boundstates of them. The open strings we consider carry angular momentum on an S$^3$ embedded 
in the S$^5$ of the AdS$_5\times$S$^5$ background. The main result of this article is that we derive 
a bosonic lattice Hamiltonian that describes the one loop mixing of the operators dual to the general 
excited giant graviton system. A semiclassical analysis of the Hamiltonian allows us to give a 
geometrical interpretation for the labeling used to describe the gauge theory operators. We also argue 
that AdS giant gravitons are unstable against the excitations considered.}

\preprint{WITS-CTP-032}

\title{Giant Gravitons - with Strings Attached (II)}


\def \Tr{\mbox{Tr\,}}

\begin{document}

\section{Introduction}

Giant gravitons\cite{McGreevy:2000cw} provide an ideal laboratory in which non-perturbative effects in string theory can be studied.
The operators dual to giant gravitons moving in the AdS$_5\times$S$^5$ background are 
known\cite{Balasubramanian:2001nh},\cite{Corley:2001zk},\cite{Berenstein:2004kk}; further, they enjoy
non-renormalization properties, so that certain
computations done at weak coupling can reliably be extrapolated to strong
coupling. This is important since we would like to compare field theory results (which we obtain at {\it weak coupling}
where we can actually do calculations) with results from the dual quantum gravity defined on a large space with small
curvature (which should reproduce the {\it strong coupling} dynamics of the quantum field theory)\cite{Maldacena:1997re}.

The giant gravitons we consider in this article correspond to ${1\over 2}$ BPS operators in the ${\cal N}=4$ super Yang-Mills theory.
These giant gravitons can be excited by attaching open strings to them. 
Operators dual to the open string plus giant graviton system were proposed in \cite{Balasubramanian:2004nb}\footnote{See 
\cite{Balasubramanian:2002sa},\cite{Strings},\cite{CuntzChain},\cite{deMelloKoch:2005jg},\cite{Berenstein:2006qk},\cite{adsgiant},\cite{strings},
\cite{Shanin} for further studies of non-BPS excitations. Some of these excitations have been interpreted as open strings attached to giant gravitons.}. 
Since the worldvolume of a giant graviton is a compact space, the Gauss law imposes strict constraints on the allowed
open string excitations. 
It is a nontrivial piece of evidence for the proposal of \cite{Balasubramanian:2004nb}, that the operators dual to the
open string plus giant system are perfectly consistent with these constraints.
Recently, a graphical notation for these operators together with the technology to compute free field theory correlation 
functions has been developed in \cite{jelena}. 
Our goal in this article is to obtain the one loop matrix of anomalous dimensions of these operators.

In a remarkable paper, Minahan and Zarembo\cite{Minahan:2002ve} showed that the spectrum of one loop anomalous dimensions
of operators dual to closed string states, in a subsector of the theory, gives rise to an integrable $SO(6)$ spin chain. 
This result was generalized to include the full set of local operators 
of the theory\cite{Beisert:2003yb}. The full planar one loop spectrum
of anomalous dimensions gives an integrable spin chain model that can 
be solved by Bethe-Ansatz techniques\cite{Beisert:2003yb}.
A similar approach for operators dual to open strings is frustrated by the fact that, since the open string 
and giant can exchange momentum, the number of sites of the open string lattice becomes a dynamical variable\footnote{An
exception to this is the case of an open string attached to a maximal giant graviton\cite{Strings}.}.
This was circumvented in \cite{CuntzChain} by introducing a Cuntz oscillator chain. 
Restricting to the $SU(2)$ sector, the spin chain is obtained by mapping one of the matrices, say $Z$, into a spin up
and the other, say $Y$, into a spin down. In contrast to this, the Cuntz chain uses the $Y$s to set up a lattice
which is populated by the $Z$s.   
Thus the number of sites in the Cuntz chain is fixed;
the fact that the open string can exchange momentum with the giant is reflected in the fact that there are sources
and sinks (at the endpoints of the string) for the particles on the chain. The precise structure of these boundary
interactions is rather complicated; indeed since the brane can exchange momentum with the string, the brane will in general
be deformed by these boundary interactions. The goal of this article is to determine this Cuntz chain Hamiltonian
for a single string attached to an arbitrary system of giant gravitons. In particular, this entails accounting for
back reaction on the giant graviton.

In section 2 we start by recalling the definition of the operators dual to a giant graviton with a single string attached.
This allows us to introduce the notation we use for Cuntz chain states. We also recall the bulk terms in the Cuntz oscillator
chain Hamiltonian that are independent of the brane system that the open string is attached to. In section 3 we describe how
to obtain the boundary interaction terms in the Hamiltonian for an arbitrary open string/brane bound state system. Section 4 
discusses the numerically tractable toy model obtained by considering a string with a single site. We come to the disappointing
conclusion that our Hamiltonian does not accurately describe the open string dynamics for this toy model. In section 5 we obtain
sigma models that describe the continuum limit of our Cuntz chains. Our results suggest that the AdS giant gravitons are unstable.
Finally, in section 6, we present our conclusions.

\section{Attaching Open Strings to Giant Gravitons}

In this section we will introduce the operators in ${\cal N}=4$ super Yang-Mills theory that are dual 
to an open string plus giant graviton system. These operators were 
originally introduced in \cite{Balasubramanian:2004nb}. 
Our goal in this article is to obtain the one loop matrix of anomalous dimensions of these operators. 
We will do this by mapping the spectrum of anomalous dimensions into a Cuntz oscillator chain model\cite{CuntzChain}.
The dynamics of the Cuntz chain has two contributions, one coming from the bulk of the string and one from the end points. The
bulk terms, which are independent of the details of the brane the open string is attached to, are known\cite{CuntzChain}. These
bulk terms are briefly reviewed in this section. The end point interactions describe how the open string interacts with the
giant it is attached to, and consequently, depends sensitively on the details of the brane state. One of the main results of this
article is the computation of these end point interactions. This is dealt with in the next section.

We study the Lorentzian ${\cal N}=4$ super Yang-Mills theory on $R\times S^3$. 
The $1/2$-BPS (and systematically small deformations of these)
states of the theory on $R\times S^3$ can be described in the
s-wave reduction of the Yang-Mills theory, i.e. in a matrix quantum mechanics \cite{Berenstein:2004kk}.
According to the state-operator correspondence 
of conformal field theory, the generator associated to dilatations on $R^4$ becomes the Hamiltonian for the theory on $R\times S^3$.
The action of ${\cal N}=4$ super Yang-Mills theory on $R\times S^3$ is
$$ S={N\over 4\pi \lambda}\int dt \int_{S^3} {d\Omega_3\over 2\pi^2}
\left( {1\over 2}(D\phi^i)(D\phi^i)+{1\over 4}\big(\big[\phi^i,\phi^j\big]\big)^2
-{1\over 2}\phi^i\phi^i +\dots \right),$$
where $\lambda =g_{YM}^2N$ is the 't Hooft coupling, $i,j=1,...,6$ and $\dots$ are the fermion and the gauge kinetic
terms in the action which we will not need here.  The mass term arises from 
conformal coupling to the metric of $S^3$. Group the six real scalars into three complex fields
$$ Z=\phi^1+i\phi^2,\qquad Y=\phi^3+i\phi^4,\qquad X=\phi^5+i\phi^6 .$$
In what follows we use these complex combinations.

\subsection{Operators dual to Excited Giants}

The dual of a giant graviton is a Schur polynomial\cite{Corley:2001zk}\footnote{In this paper we study the theory with gauge group
$U(N)$. For the extension to gauge group $SU(N)$, one needs to account for the fact that the $Z$s in this case are traceless. See
\cite{deMelloKoch:2004ws} for further details.}
\begin{equation}
\chi_R (Z)={1\over n!}\sum_{\sigma\in S_n}\chi_R (\sigma )\Tr (\sigma Z^{\otimes n}),
\label{Schur}
\end{equation}
$$\Tr (\sigma Z^{\otimes n}) =Z^{i_1}_{i_{\sigma (1)}}Z^{i_2}_{i_{\sigma (2)}}\cdots
Z^{i_{n-1}}_{i_{\sigma (n-1)}}Z^{i_n}_{i_{\sigma (n)}}.$$
Schur polynomials are labeled by Young diagrams, denoted $R$ above. A Schur
polynomial labeled by a Young diagram with a single column of length $O(N)$ is dual to a sphere giant\cite{Balasubramanian:2001nh};
a Schur polynomial labeled by a Young diagram with a single row of length $O(N)$ is dual to an AdS
giant\cite{Corley:2001zk},\cite{adsgiant}. It is natural to guess that a Schur polynomial labeled by a Young diagram with $O(1)$
columns and $O(N)$ rows is dual to a bound state of sphere giants and that
a Schur polynomial labeled by a Young diagram with $O(N)$
columns and $O(1)$ rows is dual to a bound state of AdS giants. 

One can excite giant gravitons by attaching open strings to them.
Each open string is described by a word, $W$, with $O(\sqrt{N})$ letters. These letters can
in principle be fermions, Higgs fields or covariant derivatives of these fields. We will consider open strings moving 
with a large angular momentum on the $S^5$, in the direction corresponding to $Y$. The number of $Y$ fields tells us 
the spacetime angular momentum of the string state. To describe strings moving with a large angular momentum on the $S^5$,
take words with $O(\sqrt{N})$ $Y$ letters in the word. To describe different string states, insert letters into this word. 
The remaining letters can be put 
into a correspondence with oscillators of the string worldsheet theory\cite{Balasubramanian:2002sa}. In this article we will
consider only the open string states obtained by inserting $Z$ Higgs fields so that the
open strings can have a component of angular momentum in the direction of the giant. 
Our labeling for the open string words is the following (there are $L+1$ $Y$s in $W$)
$$ (W(\{n_1,n_2,\cdots , n_L\}))^i_j= (YZ^{n_1}YZ^{n_2} Y\cdots YZ^{n_L}Y)^i_j.$$
Geometrically we are thinking of the $Y$'s as forming a lattice that is populated with $Z$'s. 
The numbers $n_i$ give the occupation number representation
of the $Z$s in this lattice. The BMN loops\cite{Berenstein:2002jq} are given by moving to momentum space on this lattice.
The endpoints of the open string are given by the first and $L+1$th $Y$ of the above word.

The proposal of \cite{Balasubramanian:2004nb} for the operators dual to excited giant gravitons
inserts the words $(W^{(a)})^j_i$ describing the open strings
(one word for each open string) into the operator describing the system of giant gravitons
\begin{equation}
\chi_{R,R_1}^{(k)}(Z,W^{(1)},...,W^{(k)})={1\over (n-k)!}
\sum_{\sigma\in S_n}\Tr_{R_1}(\Gamma_R(\sigma))\Tr(\sigma Z^{\otimes n-k}W^{(1)}\cdots W^{(k)}),
\label{restrictedschur}
\end{equation}
$$\Tr (\sigma Z^{\otimes n-k}W^{(1)}\cdots W^{(k)})= Z^{i_1}_{i_{\sigma (1)}}Z^{i_2}_{i_{\sigma (2)}}\cdots
Z^{i_{n-k}}_{i_{\sigma (n-k)}}(W^{(1)})^{i_{n-k+1}}_{i_{\sigma (n-k+1)}}\cdots
(W^{(k)})^{i_{n}}_{i_{\sigma (n)}}.$$
The label $R$ of the giant graviton system is a Young diagram with $n$ boxes, i.e. it also labels a representation 
of the symmetric group $S_n$.
$\Gamma_R(\sigma )$ is the matrix representing $\sigma$ in irreducible representation $R$ of $S_n$.
The representation $R_1$ is a Young Diagram with $n-k$ boxes, i.e. it labels a representation of $S_{n-k}$. 
By taking an $S_{n-k}$ subgroup of $S_n$ (there are many different ways to get this subgroup - see 
\cite{Balasubramanian:2004nb},\cite{jelena}), $R_1$ will be one of the representations subduced.
$\Tr_{R_1} (\cdot)$ is an instruction to trace only over the subduced $R_1$ subspace.
In \cite{jelena} this 
operator was called a restricted Schur polynomial of representation $R$ with $R_1$ the representation of the restriction.
The number of boxes in $R_1$ gives the number of $Z$'s in the giant system. Further details of the construction of this operator are not
needed in this article. We refer the interested reader to \cite{jelena} for additional details.
In this article we consider the case of a single string, that is, $k=1$.
\begin{figure}[t]\label{fig:cgraph1}
\begin{center}
\includegraphics[height=9cm,width=12cm]{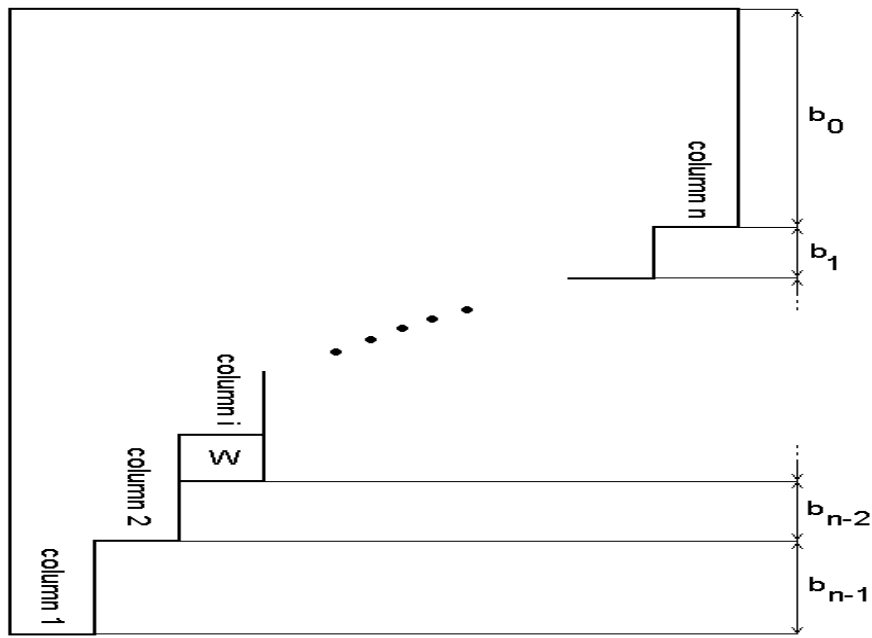}
\caption{The Young diagram shown labels an excited bound state of sphere giants. There is a single open string attached to
column $i$.} 
\end{center}
\end{figure}

We will use $K$ to denote the total number of $Z$ fields in the operator $\chi_{R,R_1}^{(1)}(Z,W)$ and $J$ to denote the number of
$Z$ fields in $W$. Thus, $R_1$ has a total of $K-J$ boxes. It is only when $J+L$ is $O(\sqrt{N})$ and $K$ is $O(N)$  
that we can interpret $\chi_{R,R_1}^{(1)}(Z,W)$ as dual to a string plus brane system. 

Since $R_1$ is obtained from $R$ by removing a single box, we have specified the operator dual to an excited giant graviton
if we have given $R$, the open string word and have stated which box is to be removed 
to obtain $R_1$. We will use the graphical notation of \cite{jelena} in which
the operator is labeled by the Young diagram $R$ itself, and the box to be removed is indicated by writing the open string word
$W$ in it. 
In figure 1 we have shown the label for a bound state of sphere giants with a single string attached. 
Later by employing the state operator correspondence of the conformal field theory, 
we will obtain a Cuntz oscillator chain. Instead of drawing this label,
we will denote the state that corresponds to this operator
by $|b_0,b_1,...,b_{n-1};W;i\rangle$. The case $n=1$ has been studied in 
detail\cite{Balasubramanian:2004nb},\cite{Balasubramanian:2002sa},\cite{Strings},\cite{CuntzChain},\cite{Berenstein:2006qk}.
We will extend the analysis to $n>1$.

In figure 2 we have shown the label for a bound state of AdS giants with a single string attached. 
After employing the state operator map to obtain
the Cuntz oscillator chain, we will replace this operator by a corresponding state.
We denote the state by $|a_0,a_1,...,a_{n-1};W;i\rangle$ instead of drawing the label.
The case $n=1$ was considered in \cite{adsgiant}. However, even for $n=1$, the analysis we perform here is different.
For our analysis, the open string word is a lattice made using the $Y$'s; we then populate this lattice with $Z$'s.
In \cite{adsgiant}, the open string word is a lattice built using covariant derivatives; again this lattice is populated
with $Z$'s. Physically our open strings have a large momentum on an $S^3$ contained in the S$^5$ while the strings of \cite{adsgiant} have a
large momentum on the $S^3$ contained in the AdS$_5$ space.
\begin{figure}[t]\label{fig:cgraph2}
\begin{center}
\includegraphics[height=9cm,width=12cm]{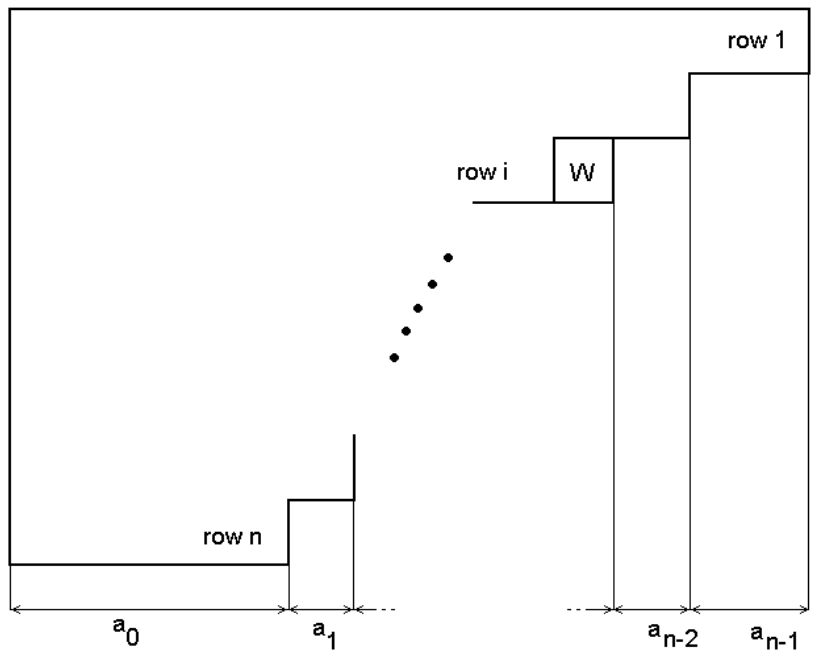}
\caption{The Young diagram shown labels a excited bound state of AdS giants. There is a single open string attached to row $i$.} 
\end{center}
\end{figure}

\subsection{Parameter Scaling}

We are interested in determining the mixing matrix of anomalous dimensions for the operators dual to excited giant gravitons with a single 
string attached. To obtain operators dual to giant gravitons, we take $b_0$ to be $O(N)$ and $b_i$ $i=1,...,n-1$ to be $O(1)$ for
the sphere giants and $a_0$ to be $O(N)$ and $a_i$ $i=1,...,n-1$ to be $O(1)$ for the AdS giants. We want to compute this mixing matrix to 
one loop and at large $N$. This is a hard problem: since the number of fields in the giant graviton is $O(N)$, the planar approximation fails.
To get an accurate result, we need to contract all of the fields in the two giant gravitons exactly. The number of fields in each word $W$ 
is $J+L\approx L$ in the case that $J\ll L$ which we assume. If the $W$ is to be dual to an open string, we need to take $L\sim O(\sqrt{N})$. 
We will not contract the open strings words exactly - only the planar diagrams are summed. To suppress the non-planar contributions, we need 
to take ${L^2\over N}\ll 1$. Concretely, we have a double scaling limit in mind, in which the first limit takes $N\to\infty$ holding 
${L^2\over N}$ fixed and then the second limit takes the effective genus counting parameter ${L^2\over N}$ to zero.
In the dual string theory, taking the limits in this way corresponds to taking the string coupling to zero, in 
the string theory constructed in a fixed giant graviton background.
Finally, we will drop contributions coming from contractions between $Z$s in the open string $W$ and $Z$s
associated to the brane system. When computing two point functions in free field theory, as long as the number of boxes in the representation 
$R$ is less than $O(N^2)$ and the numbers of $Z$'s in the open string is $O(1)$, the contractions between any $Z$s in the open string and the 
rest of the operator are suppressed in the large $N$ limit\cite{recent}. To ensure that the number of boxes in the representation 
$R$ is less than $O(N^2)$, we also assume that $n$ is $O(1)$.

Other interesting parameters to consider are $N-b_0$ and $N+a_0$. The parameter $N-b_0$ can scale as $O(N)$, $O(\sqrt{N})$ or $O(1)$.
We will see, from the results of section 3, that when $N-b_0$ is $O(1)$ the sphere giant boundary interaction is $O({1\over N})$,
when $N-b_0$ is $O(\sqrt{N})$ the boundary interaction is $O({1\over\sqrt{N}})$ and when $N-b_0$ is $O(N)$, the boundary interaction
is $O(1)$. Since we are interested in the dynamics arising from the boundary interaction, we will assume that $N-b_0$ is $O(N)$.
The boundary interaction is always $O(1)$ for the AdS giants because $a_0+N$ is always $O(N)$.

Our analysis is only valid if $J$ is $O(1)$. Cases in which $J$ becomes large correspond to the situation in which a lot of momentum is 
transferred from the giant to the open string, presumably signaling an instability. The value of $J$ is not a parameter that we can choose; 
it is determined by the dynamics of the problem. In what follows, we solve for the value of $J$. In cases where it turns out to be $O(1)$,
it can be dropped and back reaction on the giant is not important. In cases where $J$ is large, back reaction is important and the 
approximations we are employing are no longer valid. The assumption that we can drop non-planar contributions when contracting the open string 
words breaks down, essentially because as more and more $Z$s hop onto the open string, it is starting to grow into a state best described 
as a giant graviton. One can also no longer neglect the contractions between any $Z$s in the open string and the rest of the operator, 
presumably because the composite system no longer looks like a string plus giant (which can be separated nicely) but rather, it looks
like one large deformed membrane.

The process in which the word $W$ ``fragments'' thereby allowing $Y$s to populate
more than a single box in $R$ corresponds to a splitting of the original string into smaller strings, which are still attached to the giant.
This process was considered in \cite{jelena}; it does not contribute in the large $N$ limit. Finally, there is also a process in which
the open string detaches from the brane system and is emitted as a closed string state, so that it no longer occupies any box in $R$. This
process also does not contribute in the large $N$ limit\cite{jelena}.

In what follows we use the results of \cite{jelena} to contract the fields in the two giant gravitons
exactly, and we contract the open string words planarly ignoring contractions between $Z$s in the open string and the 
rest of the operator.

\subsection{Cuntz Chain Model}

As usual, we can decompose the potential for the scalars into D terms and F terms. The advantage of this
decomposition is that for the operators we study here, it is known that at one loop,
the D term contributions cancel with the
gauge boson exchange and the scalar self energies\cite{Constable:2002hw}. Consequently we will only 
consider the planar interactions arising from the F term. 

For any conformal field theory, we can trade our (local) operators for a set of states. Concretely, this involves
quantizing with respect to radial time. Considering a fixed ``radial time" slice we obtain a round sphere. In this process,
we trade the conformal dimension of our operator for the energy of the corresponding state. 
As discussed above, we interpret the $Y$ fields
in the operator as a ``lattice" which can be populated by inserting impurities (in this case $Z$'s) into the lattice
(between the $Y$'s). The F term interaction preserves the number of $Y$'s (the lattice is not dynamical) and
allows impurities to hop between neighboring sites. This interpretation thus maps the problem of determining the
anomalous dimensions of operators in the super Yang-Mills theory into the dynamics of a 
Cuntz oscillator chain. The bulk interactions are described by the Hamiltonian
\begin{equation}
H_{bulk} = 2\lambda\sum_{l=1}^L \hat{a}_l^\dagger \hat{a}_l -\lambda\sum_{l=1}^{L-1}(\hat{a}_l^\dagger \hat{a}_{l+1}
+\hat{a}_l \hat{a}^\dagger_{l+1}),
\label{bulk}
\end{equation}
where
$$ \hat{a}_i \hat{a}_i^\dagger = I,\qquad \hat{a}^\dagger_i \hat{a}_i = I-|0\rangle\langle 0|.$$
We will not rederive this Hamiltonian.
The interested reader is referred to \cite{Berenstein:2006qk} for the details of this derivation.
The first term in the Hamiltonian tells us that each occupied site contributes $2\lambda$ to the energy. 
Notice that this contribution is independent of the number of impurities occupying the state, which is a direct
consequence of the fact that we only sum planar contractions.
This is accounted for by assigning Cuntz oscillators to the impurities, 
not the standard bosonic oscillators. The next two terms are hopping terms allowing the 
impurities to move between sites. Evidently, delocalized impurities 
lower the energy. To obtain the full Hamiltonian, we
need to include the boundary interactions arising from the string/brane 
system interaction. This interaction, which introduces
sources and sinks for the impurities at the boundaries 
of the lattice, is derived in the next section.

\section{Boundary Interactions}

One of the interactions we can consider allows a $Z$ to hop from the first or last site of the string onto the giant, or from the
giant into the first or last site of the string. In the process the string exchanges momentum with the giant graviton. In addition
to these momentum exchanging processes, there is also a boundary interaction in which a $Z$ belonging to the giant ``kisses"
the first (or last) $Y$ in the open string word so that no momentum is exchanged. 
Using the formula derived in appendix A we will be able to derive the term
in the Hamiltonian describing the ``hop off" process, in which a $Z$ hops off the string and onto the giant. Since the Hamiltonian
must be Hermitian, we can obtain the ``hop on" term by daggering the ``hop off" term. 
We obtain the momentum conserving boundary interaction by expressing the kiss as a hop on followed by
a hop off. We end this section by summarizing our result for the Cuntz oscillator chain Hamiltonian.

\subsection{Hop Off Rules}

We will start by deriving the hop off interaction, for the case that the open string is attached to a single sphere giant 
or a single AdS giant. This will serve both to illustrate our method and further, to show that we recover the known boundary
interaction in this case. We will then generalize to bound states of giant gravitons. This allows us to determine
the general structure of the hop off interaction. 

\subsubsection{Single Giant Graviton}

The hopping interaction allows impurities to hop off the string and onto the giant. 
Concretely, this hop takes
$$ W(\{n_1,n_2,\cdots,n_L\})\to ZW(\{n_1-1,n_2,\cdots,n_L\})\quad {\rm or}$$
$$ W(\{n_1,n_2,\cdots,n_L\})\to W(\{n_1,n_2,\cdots,n_L-1\})Z .$$
To determine the corresponding term in the interaction Hamiltonian, we need to be able to express
objects like $\chi_{R,R_1}^{(1)}(Z,ZW)$ in terms of $\chi_{S,S_1}^{(1)}(Z,W)$. Using the formulas
in appendix A, we have
$$\chi_{\young({\,},{\,},{\,},{\,},w)}-\chi_{\young({\,},{\,},{\,},{\,})}\Tr (w)=
-\chi_{\young({\,},{\,},{\,},v)},\qquad v=Zw ,$$
$$\chi_{\young({\,}{\,}{\,}{\,}w)}-\chi_{\young({\,}{\,}{\,}{\,})}\Tr (w)=
\chi_{\young({\,}{\,}{\,}v)},\qquad v=Zw .$$
Using $(1^{b_0})$ to denote the Young diagram with a single column of $b_0$ boxes\footnote{Thus, $(1^0)\ne (1^1)$. $(1^0)$ is the
diagram with no boxes; $(1^1)$ has one box.} and $(a_0)$
to denote the Young diagram with a single row of $a_0$ boxes, the above relations can be rewritten,
in general, as
\begin{equation}
\label{sphereidentity}
\chi_{(1^{b_0+1}),(1^{b_0})}^{(1)}(Z,W)-\chi_{(1^{b_0})}(Z)\Tr (W)=-\chi_{(1^{b_0}),(1^{b_0-1})}^{(1)}(Z,ZW),
\end{equation}
\begin{equation}
\label{adsidentity}
\chi_{(a_0+1),(a_0)}^{(1)}(Z,W)-\chi_{(a_0)}(Z)\Tr (W)=
\chi_{(a_0),(a_0-1)}^{(1)}(Z,ZW) .
\end{equation}
We would like to rewrite these statements in terms of the states of the Cuntz oscillator chain.
It is convenient to normalize the states of the Cuntz oscillator chain. Normalized states
correspond to operators whose two point function is normalized. Using the technology of \cite{jelena}
it is a simple task to compute the equal time correlator. Using the propagators
$$ \langle Z^\dagger_{ij}(t)Z_{kl}(t)\rangle = {4\pi\lambda\over N}\delta_{il}\delta_{jk} = 
\langle Y^\dagger_{ij}(t)Y_{kl}(t)\rangle ,$$
we obtain
$$\langle (\chi_{(1^{b_0+1}),(1^{b_0})}^{(1)}(Z,W))^\dagger \chi_{(1^{b_0'+1}),(1^{b_0'})}^{(1)}(Z,W')\rangle
=$$
\begin{equation}
\left({4\pi\lambda\over N}\right)^{b_0+h}
\delta_{b_0 b_0'}\delta_{WW'}N^{h-1}(b_0+1){N!\over (N-b_0-1)!},
\label{giantoverlap}
\end{equation}
where we have used $h=J+L$ to denote the number of fields in $W$.
This is not the exact result for the two point function. In the language of \cite{jelena}, the $F_1$ open string
contraction has been dropped. Relative to the leading term, the dropped term is of order (this is an upper bound
for the dropped term, obtained by assuming that the word is made only out of one type of field; if there are both
$Z$s and $Y$s in $W$, this number is typically reduced by a factor of $h$) 
$$ {h(N-b_0)\over N(b_0+1)}. $$
Further, we have only summed
the planar diagrams when contracting $W$ and $W'$. If any of the $h$ fields in $W$ are $Z$s, we can have
contractions between these fields and the $Z$ fields appearing in the giant. These contractions have been dropped.
When computing two point functions in free field theory, as long as the number of boxes in the representation $R$ 
is less than $O(N^2)$ and the numbers of $Z$'s in the open string is $O(1)$, the 
contractions between any $Z$s in the open string and the rest of the operator are suppressed in the large $N$ limit\cite{recent}.
The delta function $\delta_{WW'}$ is one if the set of occupation
numbers of the two open strings are equal and is zero otherwise. Next, consider (this is again an upper bound
obtained by assuming that the word $W$ is made only out of one type of field)
\begin{equation}
\langle (\chi_{(1^{b_0})}(Z)\Tr(W))^\dagger \chi_{(1^{b_0'})}(Z)\Tr(W')\rangle
=\left({4\pi\lambda\over N}\right)^{b_0+h}
\delta_{b_0 b_0'}\delta_{WW'}h N^{h}{N!\over (N-b_0)!}.
\label{closedgiant}
\end{equation}
Compare (\ref{closedgiant}) to (\ref{giantoverlap})
\begin{equation} 
{h N^{h}{N!\over (N-b_0)!}\over N^{h-1}(b_0+1){N!\over (N-b_0-1)!}}=
{N h\over (b_0+1)(N-b_0)}.
\label{sublead}
\end{equation}
This is clearly subleading in our case where $b_0\sim O(N)$, $N-b_0\sim O(N)$ and $h\sim O(\sqrt{N})$. 
In this regime of the parameters the subleading term is naturally interpreted as a state containing a giant
graviton and a closed string. The fact that these closed string contributions are subleading 
and hence do not contribute in the leading order is a general conclusion valid in all of the situations
we consider in this article. The correspondence between operators and (normalized) states of the Cuntz oscillator
chain is
$$ \chi_{(1^{b_0+1}),(1^{b_0})}^{(1)}(Z,W))\leftrightarrow
\sqrt{\left({4\pi\lambda\over N}\right)^{b_0+h}N^{h-1}(b_0+1){N!\over (N-b_0-1)!}}|b_0+1; W; 1\rangle . $$
The hop off interaction acts as
$$ H|b_0+1; W(\{n_1,n_2,\cdots,n_L\});1\rangle \to |b_0+1; ZW(\{n_1-1,n_2,\cdots,n_L\});1\rangle .$$
After dropping the closed string
contributions, writing things in terms of the states of the Cuntz oscillator chain, and employing
(\ref{sphereidentity}) we obtain
(we want to consider the hop off process for a giant with momentum $b_0$ and hence we start with a single
column containing $b_0+1$ boxes; this is {\it not} the complete hop off interaction - we have only shown the term
obtained when a $Z$ hops out of the the first site)
\begin{eqnarray}
\nonumber 
H|b_0+1; W(\{n_1,n_2,\cdots,n_L\});1\rangle &=& -\sqrt{1-{b_0\over N}}|b_0+1; ZW(\{n_1-1,n_2,\cdots,n_L\});1\rangle\\
\label{boundaryrule}
=-\sqrt{1-{b_0\over N}}|b_0&+&2;W(\{n_1-1,n_2,\cdots,n_L\});1\rangle.
\end{eqnarray}
If we introduce the operator $\hat{A}^\dagger$ that increases the number of $Z$s in the giant by 1, the
boundary hop off interaction can be written as (this term in the Hamiltonian is positive because the 
piece of the F term that generates this interaction is negative and we have a $-$ sign in our rule
(\ref{boundaryrule}))
\begin{equation}
H=\lambda\sqrt{1-{b_0\over N}} \hat{A}^\dagger \hat{a}_1 .
\label{hopoffsphere}
\end{equation}
This interaction Hamiltonian vanishes for the maximal giant\cite{Strings} and is highly suppressed for giants which are close
to maximal\cite{CuntzChain}.
Notice that the interaction is not proportional to the number of $Z$'s in the giant. 
For this reason, we choose the oscillator $\hat{A}^\dagger$ to be a Cuntz oscillator
$$ \hat{A}\hat{A}^\dagger = I,\qquad \hat{A}^\dagger \hat{A} = I-|0\rangle\langle 0|.$$
Since the total number of $Z$s in the operator is conserved, $b_0$ is the difference between the total number of $Z$s ($=K$) and
the number of $Z$s on the string. This gives the expression
$$ b_0=K-\sum_{n=1}^{\infty}\sum_{l=1}^L (\hat{a}_l^\dagger)^n (\hat{a}_l)^n .$$
Finally, since the impurity can either hop out of the first or the last sites, we can write
the complete hop off interaction, for a string attached to a single sphere giant, as
$$H=\lambda \sqrt{1-{K-\sum_{n=1}^{\infty}\sum_{l=1}^L (\hat{a}_l^\dagger)^n (\hat{a}_l)^n-1\over N}} 
(\hat{A}^\dagger \hat{a}_1 + \hat{A}^\dagger \hat{a}_L) .$$

For the AdS giants, the relevant two point functions are
$$\langle (\chi_{(a_0+1),(a_0)}^{(1)}(Z,W))^\dagger \chi_{(a'_0+1),(a'_0)}^{(1)}(Z,W')\rangle
=\delta_{a_0 a_0'}\delta_{WW'}N^{h-1}(a_0+1){(N+a_0)!\over N!}\left(
{4\pi\lambda\over N}\right)^{a_0+h},$$
$$\langle (\chi_{(a_0)}(Z)\Tr (W))^\dagger \chi_{(a'_0)}(Z)\Tr (W')\rangle
=\delta_{a_0 a_0'}\delta_{WW'} h N^h {(N+a_0-1)!\over N!}\left(
{4\pi\lambda\over N}\right)^{a_0+h}.$$
In the first correlator we have again dropped the $F_1$ open string contraction; relative to the leading term it is of order
(this is again using an upper bound
for the dropped term, obtained by assuming that the open string word is made only out of one type of field)
$$ {h(N+a_0)\over N(a_0+1)},$$
which is subleading for $h\sim O(\sqrt{N})$ and $a_0\sim O(N)$. Looking at the second correlator we again conclude that
the closed string contribution is subleading. Writing (\ref{adsidentity}) in terms of states of a Cuntz oscillator
chain state, we obtain (we want to consider the hop off process for a giant with momentum $a_0$ and
hence start with a single row containing $a_0+1$ boxes; this term is obtained if $Z$ hops out of the first site)
$$ H|a_0+1; W(\{n_1+1,n_2,\cdots,n_L\}); 1\rangle =\sqrt{1+{a_0\over N}}|a_0+2;W(\{n_1,n_2,\cdots,n_L\}); 1\rangle .$$
Again, it is a simple matter to modify the above argument to prove that (this term is obtained if $Z$ hops out of the last site)
$$ H|a_0+1; W(\{n_1,n_2,\cdots,n_L+1\}) ; 1\rangle =\sqrt{1+{a_0\over N}}|a_0+2;W(\{n_1,n_2,\cdots,n_L\}); 1\rangle .$$
Using these identities, we find that the hop off interaction, for a string attached to a single AdS giant, is
\begin{equation}
H= -\lambda\sqrt{1+{K-\sum_{n=1}^{\infty}\sum_{l=1}^L 
(\hat{a}_l^\dagger)^n (\hat{a}_l)^n-1\over N}} (\hat{A}^\dagger \hat{a}_1 + \hat{A}^\dagger \hat{a}_L) .
\label{hopoffads}
\end{equation}
Notice that now the interaction is enhanced as the momentum of the giant grows, in contrast to the sphere giant.
This structure of the boundary interaction was also obtained in \cite{adsgiant}, where the $Z$s hop on a lattice made
from covariant derivatives. The relative sign difference between (\ref{hopoffsphere}) and (\ref{hopoffads}) is not meaningful;
it can be eliminated, for example, by redefining the phases of the sphere giant states.

\subsubsection{Boundstate of Giants}

The first boundstate we will consider is a boundstate of two sphere giants. A Young diagram with $b_0+b_1$ boxes in the first
column and $b_0$ boxes in the second column will be denoted as $(2^{b_0}1^{b_1})$. Then, using the formula derived in appendix
A, a little algebra shows that
$$\chi^{(1)}_{(2^{b_0}1^{b_1}),(2^{b_0}1^{b_1-1})}(Z,ZW) =
-{b_1(b_1+2)\over (b_1+1)^2}\left[
\chi^{(1)}_{(2^{b_0}1^{b_1+1}),(2^{b_0}1^{b_1})}(Z,W)
-\chi_{(2^{b_0}1^{b_1})}(Z)\Tr(W)\right]$$
\begin{equation}
\label{twobound}
+{b_1\over (b_1+1)^2}\left[
\chi^{(1)}_{(2^{b_0+1}1^{b_1-1}),(2^{b_0}1^{b_1})}(Z,W)
-\chi_{(2^{b_0}1^{b_1})}(Z)\Tr(W)\right],
\end{equation}
$$\chi^{(1)}_{(2^{b_0}1^{b_1}),(2^{b_0-1}1^{b_1+1})}(Z,ZW) = 
-{b_1+2\over (b_1+1)^2}\left[
\chi^{(1)}_{(2^{b_0}1^{b_1+1}),(2^{b_0}1^{b_1})}(Z,W)
-\chi_{(2^{b_0}1^{b_1})}(Z)\Tr(W)\right]$$
\begin{equation}
\label{twoboundagain}
-{b_1(b_1+2)\over (b_1+1)^2}\left[
\chi^{(1)}_{(2^{b_0+1}1^{b_1-1}),(2^{b_0}1^{b_1})}(Z,W)
-\chi_{(2^{b_0}1^{b_1})}(Z)\Tr(W)\right].
\end{equation}
For the limit that we consider, $(N-b_0-b_1)=O(N)$, $b_0=O(N)$ and $h=O(\sqrt{N})$, where 
we again use $h$ to denote the total number of fields in $W$.
In this case, we again find that the closed string contributions are
not important in the leading order and can be dropped.
To interpret these formulas it is useful rewrite them, for
some particular values, employing our graphical notation.
If $b_1=0$, the string can only be attached to the second column.
In this case, the right hand side of (\ref{twobound}) vanishes. This is as expected, since the left hand side
would correspond to the case that $b_1=0$ and we attached the string to the first column, which is not an allowed
state\footnote{Its not allowed because if you remove the open string you are not left with a valid Young diagram, i.e.
for this state $R_1$ in (\ref{restrictedschur}) is not a valid label.}. 
Looking at (\ref{twoboundagain}), we see that the only surviving term on the right hand side corresponds to the case that the
open string is attached to the first column
$$\chi_{\young({\,}{\,},{\,}{\,},{\,}{\,},{\,}{x})}\to 
\chi_{\young({\,}{\,},{\,}{\,},{\,}{\,},{\,}{\,},{w})},\qquad x=Zw.$$
There is a subleading term (suppressed because $b_0$ is $O(N)$) which has been dropped. It has the form
\begin{equation}
\chi_{\young({\,}{\,},{\,}{\,},{\,}{\,},{\,}{x})}\to 
\chi_{\young({\,}{\,}{w},{\,}{\,},{\,}{\,},{\,}{\,})},\qquad x=Zw.
\label{dropped}
\end{equation}
The free fermion state corresponding to a Young diagram with the above shape, in the case that the number of
rows is $O(N)$ and the number of columns is $O(1)$, would contain one
fermion just above the Fermi surface and two holes deep in the Fermi sea. Thus,
the interpretation of the right hand side of (\ref{dropped}) is in terms of a bound state of sphere giants
together with a closed string (graviton) excitation\cite{Berenstein:2004kk},\cite{Lin:2004nb}. The fact that it is
$O({1\over b_0})=O({1\over N})$ is expected because the closed string coupling constant is ${1\over N}$.

In view of this example, we find a natural interpretation for the coefficients
$$C^1_{b_1} = {b_1 (b_1+2)\over (b_1+1)^2},$$
appearing in (\ref{twobound}) and (\ref{twoboundagain}).
These coefficients switch the interactions off gracefully. Indeed, $C^1_{b1}$ vanishes when $b_1$ vanishes, but very rapidly
approaches 1 as $b_1$ is increased. These coefficients multiply the terms for which the open string
remains attached to the same giant graviton. As $b_1$ is increased, the remaining coefficients in (\ref{twobound}) and
(\ref{twoboundagain}) rapidly approach ${1\over 1+b_1}$. For these terms, the string swaps from one giant to the other.
Interpret the number of boxes separating the box that the string starts in from the box that the strings lands up in, as we 
move on the right hand side of the Young diagram, as a distance. This distance is $r=1+b_1$, so that the term in which
the string swaps the giant it is attached to is essentially a ${1\over r}$ interaction.
The brane worldvolume theory describing the dynamics of the open strings attached to these giants is expected to be a
$3+1$ dimensional emergent Yang-Mills theory\cite{Balasubramanian:2001dx},\cite{Balasubramanian:2004nb}.
The ${1\over r}$ potential, which would arise from the exchange of massless particles 
in $3+1$ dimensions, thus looks rather natural.
In this article, we will call the limit in which we see these nice simplifications, the effective field theory limit.
This distance $r$ is related to the radial coordinate of the two dimensional $y=0$ plane on which the LLM boundary conditions are
specified\cite{Lin:2004nb}.
Both this distance and the ${1\over r}$ interaction were already visible in\cite{jelena}.

Finally, note that in the $b_1\to 0$ limit, the two sphere giants carry exactly the same momentum. Since their momenta
determine their radius, in this limit the two brane worldvolumes become coincident. 
Thus the $C^1_{b_1}$ coefficient is switching off a short distance membrane interaction.

We now want to write the boundary interaction term that acts on the Cuntz chain states corresponding
to normalized operators. The two point functions we need to evaluate are
$$\langle (\chi^{(1)}_{(2^{b_0}1^{b_1}),(2^{b_0}1^{b_1-1})}(Z,W))^\dagger
\chi^{(1)}_{(2^{b'_0}1^{b'_1}),(2^{b'_0}1^{b'_1-1})}(Z,W')\rangle =$$
$$ \delta_{b_0 b'_0}\delta_{b_1 b'_1}\delta_{WW'}N^{h-1}{b_1 b_0\over b_1+1}
{N!(N+1)!\over (N-b_0-b_1)!(N-b_0+1)!}
\left({4\pi\lambda\over N}\right)^{2b_0+b_1+h-1},$$
$$\langle (\chi^{(1)}_{(2^{b_0}1^{b_1}),(2^{b_0-1}1^{b_1+1})}(Z,W))^\dagger
\chi^{(1)}_{(2^{b'_0}1^{b'_1}),(2^{b'_0-1}1^{b'_1+1})}(Z,W')\rangle =$$
$$\delta_{b_0 b'_0}\delta_{b_1 b'_1}\delta_{WW'}N^{h-1}{(b_1+2) b_0\over b_1+1}
{N!(N+1)!\over (N-b_0-b_1)!(N-b_0+1)!}
\left({4\pi\lambda\over N}\right)^{2b_0+b_1+h-1},$$
$$\langle (\chi^{(1)}_{(2^{b_0}1^{b_1}),(2^{b_0}1^{b_1-1})}(Z,W))^\dagger
\chi^{(1)}_{(2^{b'_0}1^{b'_1}),(2^{b'_0-1}1^{b'_1+1})}(Z,W')\rangle = 0.$$
The $F_1$ contraction which has again been dropped, is subleading; to verify this, recall that in the limit
we consider $b_0=O(N)$, $b_1=O(1)$, $N-b_0-b_1=O(N)$ and $h=O(\sqrt{N})$.
We can now write down the action of the hop off interaction on the Cuntz chain states. To write
this interaction, again introduce the Cuntz oscillators $\hat{a}_l$ and $\hat{a}_l^\dagger$ for impurities on the string.
It is tempting to introduce a pair of Cuntz oscillators, one for each giant graviton. 
We have not employed this description. To motivate why we have used a different approach,
let $\hat{A}_i$ denote the operator that will remove a box from
the $i$th column and $\hat{A}_i^\dagger$ the operator that will insert a box into the $i$th column. 
Thus, for example, we have
$$ \hat{A}_1\,\,{}_{\yng(2,2,2,1,1)}={}_{\yng(2,2,2,1)},\qquad 
\hat{A}_2^\dagger\,\, {}_{\yng(2,2,2,1,1)}={}_{\yng(2,2,2,2,1)}.$$
When these giant oscillators act on a Young diagram, they must produce another Young diagram. 
This requirement implies that, for example
$$ \hat{A}_1\,\,{}_{\yng(2,2,2)}=0,\qquad \hat{A}_2^\dagger\,\, {}_{\yng(2,2,2)}=0.$$
Relations like these can be used to show that the oscillators for the the two giants {\it do not commute}. 
Indeed, to see that $\hat{A}_1^\dagger$ and $\hat{A}_2^\dagger$ can't commute note that
$$ \hat{A}_2^\dagger \hat{A}_1^\dagger \,\, {}_{\yng(2,2)}= {}_{\yng(2,2,2)},\,\,\,\,\,\qquad {\rm but}\,\,\,\,\,\qquad
\hat{A}_1^\dagger \hat{A}_2^\dagger \,\, {}_{\yng(2,2)}= 0.$$

Due to these complications, we have pursued an alternative description of the giants. Our alternative description
involves associating a one dimensional lattice to each Young diagram. 
Our lattice has a total of $N$ sites; each site is occupied by an arbitrary number of particles. 
If the Young diagram has $O(1)$ columns that each have $O(N)$ rows (a bound state of sphere giants),
the number of particles in the lattice is equal to the number of sphere giants in the boundstate.
We will refer to this lattice as the giant lattice to distinguish it from the string lattice.
The translation between the Young diagram and the giant lattice, is given by setting the occupation number of
lattice site $i$
$$ n_i = r_i - r_{i+1},\qquad i=1,\dots ,N,$$
where $r_i$ is the number of boxes in the $i$th row of the Young diagram
and we set $r_{N+1}=0$. There is one marked site - the site that the open string occupies. The marked site is
indicated by writing a bar above the occupation number. Two examples to illustrate the lattice notation
$$ \young({\,}{\,}{\,},{\,}{w},{\,})\leftrightarrow \{ n_1= 1, n_2=\bar{1}, n_3=1\}\qquad
   \young({\,}{\,}{\,}{\,}{\,},{\,}{w})\leftrightarrow \{ n_1= 3, n_2=\bar{2} \} . $$
We will label kets of the giant lattice by their occupation numbers. Occupation numbers that are equal to zero
are not displayed. 
The giant lattice notation is convenient because adding and subtracting boxes from the diagram has a very natural 
interpretation: adding or subtracting boxes in the first row adds or subtracts particles from the lattice. 
Adding or subtracting boxes to any other row does not change the particle 
number - its described by particles hopping on the lattice. 
When we add a box, particles hop from the $i$th to the $i+1$th site;
when we remove a box, particles hop from the $i$th to the $i-1$th site.
To describe the giant lattice, we can again introduce Cuntz oscillators - $\hat{A}_i$ and $\hat{A}^\dagger_i$, $i=1,...,N$ - one 
for each site of the giant lattice. 
Our original description of single sphere giants and AdS giants is easily translated into this new language: the dynamics
for the AdS giant is essentially single site dynamics - only the first site participates; it has occupation number
$n_1=a_0$. The dynamics is single particle dynamics for the sphere giant - the particle occupies the $b_0$th site.
Apart from the Cuntz oscillators of the giant and string lattices, we will introduce a Cuntz oscillator for the open string itself,
denoted $\hat{W}_i$ and $\hat{W}_i^\dagger$. This extra oscillator is needed to keep track of the position of the string.
In terms of these oscillators, we obtain an alternative representation of the above states. For example
$$ \young({\,}{\,}{\,},{\,}{w},{\,})\leftrightarrow \{ n_1= 1, n_2=\bar{1}, n_3=1\}
\leftrightarrow \hat{A}_1^\dagger \hat{W}_2^\dagger \hat{A}_3^\dagger |0\rangle .$$

Notice that the occupied states coincide with the position of the corners of the Young diagram.
We could also have constructed a giant lattice using the number of boxes in each column.
The reason why we have chosen to use the rows instead, is simply that the number of rows of the Young diagram
is bounded by $N$.
If we use the columns there is no such bound; further, if we try to use only the ``occupied columns" we obtain a description with a dynamical lattice.
The number of particles hopping on this dynamical lattice is equal to the number of AdS giants. 
From this point of view, the dynamical lattice appears to be the natural description for AdS giants. See appendix C for a description
of the AdS giants using the dynamical lattice formulation.
Notice that the total number of particles hopping on this dynamical lattice is constrained to be less than or equal to $N$.

It is perhaps useful to comment on why the giant lattice provides a good description.
The difficulty with introducing a pair of Cuntz oscillators - one for each column - stems from the fact that we need to impose a constraint
forcing the number of particles created by the first oscillator to be greater than or equal to the number of particles created by the second
oscillator. Indeed, occupation number states that don't satisfy this constraint would correspond to diagrams with more boxes in the second
column than in the first column - this is not a legal Young diagram. With the new giant lattice description, {\it any} occupation number
assignment leads to a valid Young diagram.

The action of the hop off interaction Hamiltonian can now be written as ($W^{(1)}=W(\{n_1,n_2,...,n_L\})$;
$W^{(2)}=W(\{n_1-1,n_2,...,n_L\})$ or $W^{(2)}=W(\{n_1,n_2,...,n_L-1\})$ if we hop off the first or last site respectively)
\footnote{In writing this contribution to the Hamiltonian, we 
have dropped ${b_1\over b_0}$ corrections, which are $O({1\over N})$ in the limit we consider. The factors ${b_0+b_1\over N}$ can be replaced
by ${b_0\over N}$; we did not make this replacement since by keeping $b_1$ it is clear that these parameters are the momenta of the two giants.}
$$ H|\{ n_{b_0}=\bar{1},n_{b_0+b_1}=1\};W^{(1)}\rangle =
\lambda\left[\sqrt{1-{b_0\over N}}
\sqrt{C_{b_1}^1}|\{ n_{b_0+1}=\bar{1},n_{b_0+b_1}=1\};W^{(2)}\rangle\right.$$
$$+\left. \sqrt{1-{b_0+b_1\over N}}{1\over b_1+1}|\{ n_{b_0}=1,n_{b_0+b_1+1}=\bar{1}\};W^{(2)}\rangle\right],$$
$$ H|\{n_{b_0}=1,n_{b_0+b_1}=\bar{1}\};W^{(1)} \rangle =\lambda\left[\sqrt{1-{b_0+b_1\over N}}
\sqrt{C^1_{b_1}}|\{n_{b_0}=1,n_{b_0+b_1+1}=\bar{1}\}; W^{(2)} \rangle\right.$$
$$-\left. \sqrt{1-{b_0\over N}}{1\over b_1+1}|\{n_{b_0+1}=\bar{1},n_{b_0+b_1}=1\};W^{(2)} \rangle\right].
$$
This result is also obtained if we consider the different limit, in which $b_1$ scales as $\sqrt{N}$. This limit is considered so that 
we can consider the situation in which the two branes are well separated in spacetime and hence when we expect 
that they stop interacting with each other. In this limit, we have
$$ H|\{ n_{b_0}=\bar{1},n_{b_0+b_1}=1\};W^{(1)}\rangle \approx
\lambda\sqrt{1-{b_0\over N}}|\{ n_{b_0+1}=\bar{1},n_{b_0+b_1}=1\};W^{(2)}\rangle,$$
$$ H|\{n_{b_0}=1,n_{b_0+b_1}=\bar{1}\};W^{(1)} \rangle \approx \lambda\sqrt{1-{b_0+b_1\over N}}
|\{n_{b_0}=1,n_{b_0+b_1+1}=\bar{1}\}; W^{(2)} \rangle ,$$
which is just two copies of the hop off interaction we found for the single giant case.

In terms of the Cuntz oscillators, the hop off interaction Hamiltonian can be written as
\begin{equation} 
H=\lambda (\hat{a}_1+\hat{a}_L)\left[
\sum_{l=1}^{N-1}\sqrt{1-{l\over N}}\hat{W}_{l+1}^\dagger\hat{W}_l\sqrt{C^1_{\hat{b}_1}}+
\sum_{l=1}^{N}\sum_{k=1}^{N-1}\sqrt{1-{k\over N}}{\epsilon (k-l)\over |k-l|+1}\hat{W}_{k+1}^\dagger
\hat{A}_k\hat{A}_l^\dagger \hat{W}_l \right.
\label{twospheregiants}
\end{equation}
$$\left. +\sum_{l=1}^{N-1}\sqrt{1-{l\over N}}\hat{W}_{l+1}^\dagger\hat{A}_l^\dagger\hat{A}_l\hat{W}_l\right],$$
where
$$ \hat{b}_1=\sum_{k=1}^{N}\sum_{l=1}^{N} |k-l|\hat{A}_k^\dagger\hat{A}_k\hat{W}_l^\dagger\hat{W}_l ,$$
and
$$ \epsilon (k)=\left\{ \matrix{-1 &{\rm if} &k<0\cr 0 &{\rm if} &k=0\cr +1 &{\rm if} &k>0}\right. .$$
This hop off interaction acts on the subspace of states of the form
$$ |\Psi\rangle = \sum_{k,l}\alpha_{kl}\hat{A}_k^\dagger\hat{W}_l^\dagger |0\rangle .$$
The hop off interaction (\ref{twospheregiants}) allows the open string to hop between rows. Note however, that the coefficients
of these hopping terms vanish for the hopping process which would allow the open string to hop into the $N+1$th row,
i.e. acting on a state which corresponds to a valid Young diagram, the Hamiltonian will produce another state which corresponds to a
valid Young diagram. 

Carrying out the same steps for a boundstate of two AdS giants we find
$$ H|\{ n_{1}=\overline{a_0+a_1},n_{2}=a_0\};W^{(1)}\rangle =
-\lambda\left[\sqrt{1+{a_0+a_1\over N}}
\sqrt{C_{a_1}^1}|\{ n_{1}=\overline{a_0+a_1+1},n_{2}=a_0 \};W^{(2)}\rangle\right.$$
$$+\left. \sqrt{1+{a_0\over N}}{1\over a_1+1}|\{ n_{1}=a_0+a_1,n_{2}=\overline{a_0+1}\};W^{(2)}\rangle\right],$$
$$ H|\{n_{1}=a_0+a_1,n_{2}=\overline{a_0} \};W^{(1)} \rangle =-\lambda\left[\sqrt{1+{a_0\over N}}
\sqrt{C^1_{a_1}}|\{n_{1}=a_0+a_1,n_{2}=\overline{a_0+1}\}; W^{(2)} \rangle\right.$$
$$-\left. \sqrt{1+{a_0+a_1\over N}}{1\over a_1+1}|\{n_{1}=
\overline{a_0+a_1+1},n_{2}=a_0\};W^{(2)} \rangle\right],
$$
for the hop off interaction. Many of the features present for a bound state of two sphere giants are present
in this result: (i) the factors $C^1_{a_1}$ gracefully turn off certain interactions as $a_1\to 0$
and this limit again corresponds to coincident membranes, (ii) the
off diagonal terms display a ${1\over r}$ dependence in the effective field theory limit 
and (iii) in the large $a_1$ limit, this contribution to the Hamiltonian
reduces to two copies of the hop off interaction for the single giant case. 

In Appendix B we give the results for boundstates of three and four giants.
From these results, it is clear that there is a general structure that can be used to write down the
hop off interaction for an arbitrary number of boundstates. Further, the features just discussed for the
boundstate of two giants hold for the general giant boundstate. If one considers the effective field theory
limit in which $b_0=O(N)$ and the $b_i$ $i=1,2,...,n-1$ are $O(N^0)$ and $\gg 1$, then the hop off term in the
Hamiltonian for $n=O(N^0)$ sphere giants takes the particularly simple form
$$ H=\lambda (\hat{a}_1+\hat{a}_L)\left[ \sum_{l=1}^{N-1}\sqrt{1-{l\over N}}\hat{W}_{l+1}^\dagger\hat{W}_l+
\sum_{l=1}^{N}\sum_{k=1}^{N-1}\sqrt{1-{k\over N}}{\epsilon (k-l)\over |k-l|}\hat{W}_{k+1}^\dagger
\hat{A}_k\hat{A}_l^\dagger \hat{W}_l \right].$$
This Hamiltonian acts on the subspace of states that have a single open string $\hat{W}^\dagger_i$ excitation and $n-1$
$\hat{A}^\dagger_j$ excitations.

\subsection{Hop On Rules}

We know that the anomalous dimensions are real. Consequently, the energies from our Cuntz chain Hamiltonian must
be real implying the Hamiltonian must be Hermitian. Thus, we can obtain the hop on term in the Hamiltonian by taking 
the Hermitian conjugate of the hop off term. As an example, the hop on term for a string attached to a single
sphere giant is given by
\begin{eqnarray}
\nonumber
\lambda\sqrt{1-{b_0\over N}}\left[
\hat{A}^\dagger\hat{a}_1 + \hat{A}^\dagger\hat{a}_L\right]^\dagger
&=& \lambda\left[
\left(\hat{A}^\dagger\hat{a}_1 + \hat{A}^\dagger\hat{a}_L\right)
\sqrt{1-{\hat{b}_0\over N}}\right]^\dagger\\
\nonumber
&=& \lambda\sqrt{1-{\hat{b}_0\over N}}\left[
\hat{A}\hat{a}_1^\dagger + \hat{A}\hat{a}_L^\dagger\right]\\
\nonumber
&=& \lambda\left[
\hat{A}\hat{a}_1^\dagger + \hat{A}\hat{a}_L^\dagger\right]
\sqrt{1-{\hat{b}_0-1\over N}}\\
\nonumber
&=& \lambda\sqrt{1-{b_0-1\over N}}\left[
\hat{A}\hat{a}_1^\dagger + \hat{A}\hat{a}_L^\dagger\right].
\end{eqnarray}
These calculations obviously assume we are working in a basis of states that have the momentum of the giant as a good
quantum number. 

For our second example, we consider a bound state of two sphere giants. A useful identity is
$$\hat{b}_1\left( \sum_{l=1}^{N-1}\sqrt{1-{l\over N}}\hat{W}_l^\dagger \hat{W}_{l+1}\right)=
\left( \sum_{l=1}^{N-1}\sqrt{1-{l\over N}}\hat{W}_l^\dagger \hat{W}_{l+1}\right)(\hat{b}_1-\hat{\epsilon}),$$
where
$$\hat{\epsilon}\equiv\sum_{l=1}^{N}\sum_{k=1}^{N}\epsilon (k-l)\hat{W}_k^\dagger\hat{W}_k\hat{A}_l^\dagger\hat{A}_l 
-\sum_{k=1}^{N} \hat{W}_k^\dagger\hat{W}_k\hat{A}_k^\dagger\hat{A}_k .$$
It is now a simple matter to verify that the hop on interaction is
$$\lambda\left[
\sum_{l=1}^{N-1}\sqrt{1-{l\over N}}\hat{W}_l^\dagger \hat{W}_{l+1}\sqrt{C^1_{\hat{b}_1-\hat{\epsilon}}}
+\sum_{l=1}^{N}\sum_{k=1}^{N-1} \sqrt{1-{k\over N}}{\epsilon (k-l)\over |k-l|+1}\hat{W}_l^\dagger\hat{A}_l
\hat{A}^\dagger_k\hat{W}_{k+1}
\right] (\hat{a}_1^\dagger +\hat{a}_L^\dagger ) .$$
\begin{figure}[t]\label{fig:cgraph2}
\begin{center}
\includegraphics[height=6cm,width=12cm]{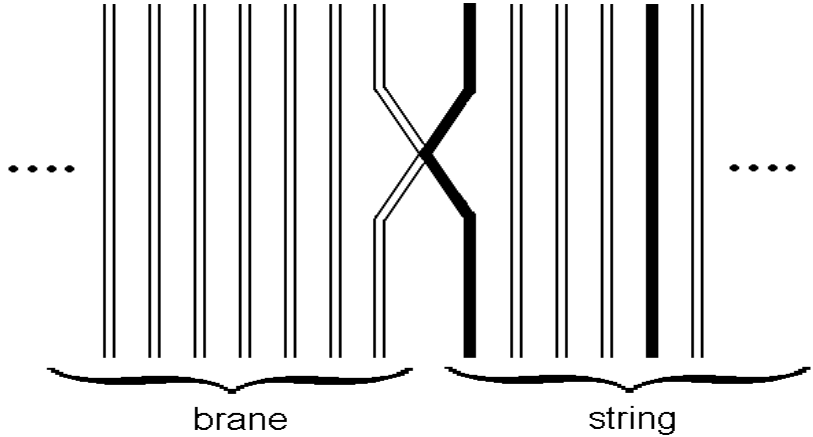}
\caption{In the Feynman diagram shown, we have an example of the kissing interaction. The white ribbons are $Z$ fields, the
black ribbons are $Y$ fields. 
The interacting black ribbon shown marks the beginning of the string; there are 3 $Z$s in the first site of the string.} 
\end{center}
\end{figure}

\subsection{Hop On and then Off for a Kiss}

The terms in our Cuntz chain Hamiltonian generate the Feynman diagrams obtained by allowing a single $F$ term vertex. 
The kissing interaction corresponds to the Feynman diagram shown in figure 3. The number of $Z$ fields in the
giant is unchanged by this process. Since the number of $Z$ fields in the giant determines the momentum of the giant,
the string and brane do not exchange momentum by this process. 
As far as the combinatorics goes, we can model the kissing interaction as a hop on (the string) followed by a hop off. Since we know both the 
hop on and hop off terms, the kissing interaction follows. Note that a hop on interaction followed by a hop off interaction will
leave the number of $Z$ fields in the giant unchanged. See figure 4. Although we have shown the diagrams using the first site of
the string for illustration, it is clear that the argument goes through for the last site as well.

\subsection{Complete Hamiltonian}
\begin{figure}[t]\label{fig:cgraph2}
\begin{center}
\includegraphics[height=8cm,width=12cm]{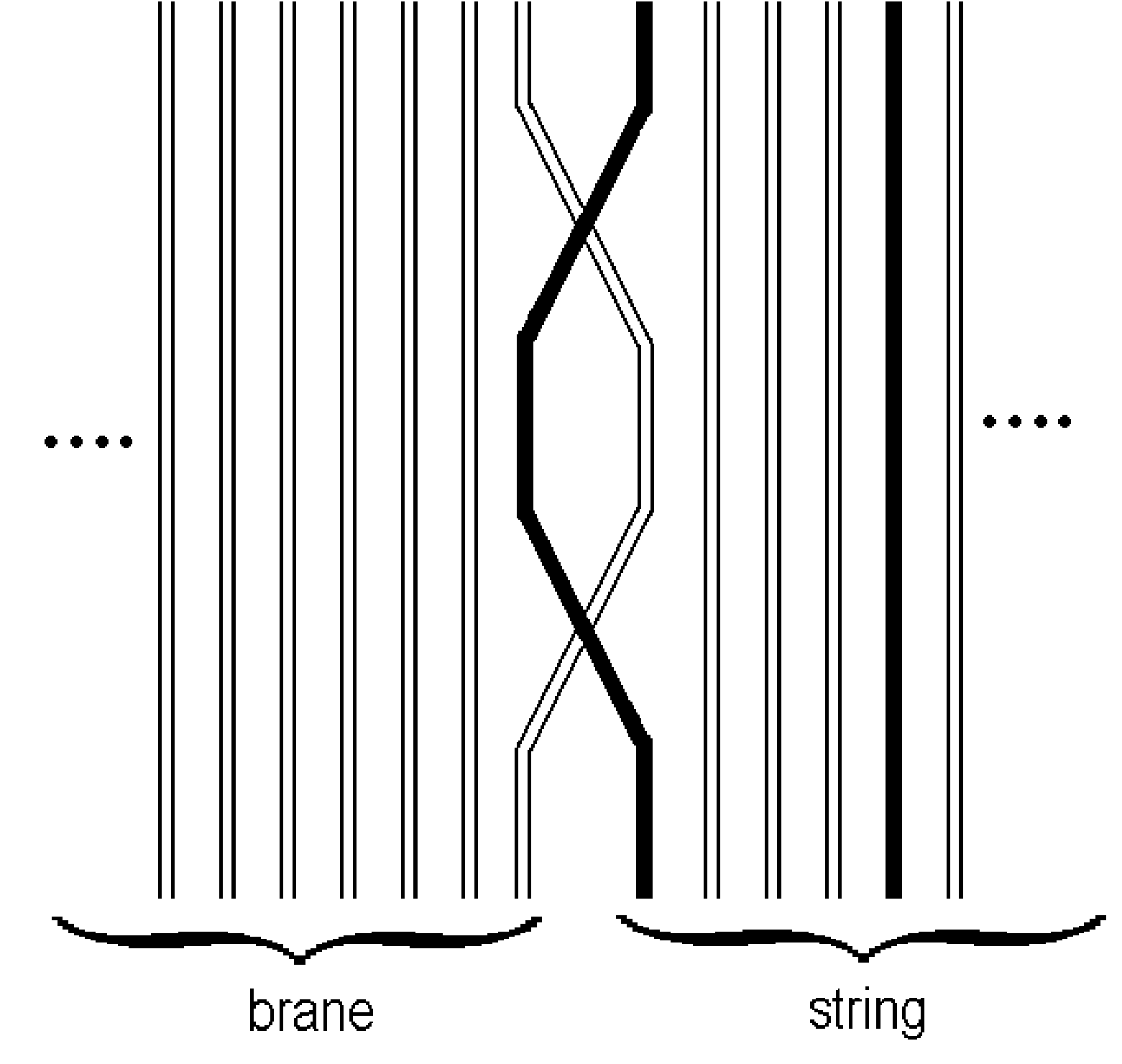}
\caption{The Feynman diagram shown has a hop on interaction followed by a hop off interaction. If you shrink the composite
hop on/hop off interaction to a point, you recover the kissing interaction.} 
\end{center}
\end{figure}

We are now in a position to assemble the complete Hamiltonian, by summing the bulk terms and the complete set of boundary interactions.
In this section we will quote the Hamiltonians we have obtained. The complete Hamiltonian is
$$ H=H_{bulk}+H_{boundary},$$
where $H_{bulk}$ is given in (\ref{bulk}). For a single sphere giant we have ($b_0$ is the momentum of the giant)
$$ H_{boundary}=\lambda \sqrt{1-{b_0\over N}} 
(\hat{A}^\dagger \hat{a}_1 + \hat{A}^\dagger \hat{a}_L)
+\lambda \sqrt{1-{b_0-1\over N}} 
(\hat{A} \hat{a}_1^\dagger + \hat{A} \hat{a}_L^\dagger )
+2\lambda\left(1-{b_0-1\over N}\right)\hat{A}^\dagger\hat{A}. $$
For a single AdS giant ($a_0$ is the momentum of the giant)
$$ H_{boundary}=-\lambda \sqrt{1+{a_0\over N}} 
(\hat{A}^\dagger \hat{a}_1 + \hat{A}^\dagger \hat{a}_L)
-\lambda \sqrt{1+{a_0-1\over N}} 
(\hat{A} \hat{a}_1^\dagger + \hat{A} \hat{a}_L^\dagger )
+2\lambda\left(1+{a_0-1\over N}\right)\hat{A}^\dagger\hat{A}.$$
We stress that these Hamiltonians have been written down assuming that we work in a basis for which
the momentum of the giant is a good quantum number. To obtain the Hamiltonian in a general basis, one can write all factors
involving the giant momenta to the right of the giant creation and annihilation operators, and then replace the 
momenta $b_i$ and $a_i$ by the corresponding number operators.

For a bound state of two sphere giants (the first column of the Young diagram has $b_0+b_1$ boxes; the second column of the Young diagram
has $b_0$ boxes)
$$ H_{boundary}=
\lambda (\hat{a}_1+\hat{a}_L)\left[
\sum_{l=1}^{N-1}\sqrt{1-{l\over N}}\sqrt{C^1_{\hat{b}_1-\hat{\epsilon}}}\hat{W}_{l+1}^\dagger\hat{W}_l+
\sum_{l=1}^{N}\sum_{k=1}^{N-1}\sqrt{1-{k\over N}}{\epsilon (k-l)\over |k-l|+1}\hat{W}_{k+1}^\dagger
\hat{A}_k\hat{A}_l^\dagger \hat{W}_l \right.
$$
$$\left. +\sum_{l=1}^{N-1}\sqrt{1-{l\over N}}\hat{W}_{l+1}^\dagger\hat{A}_l^\dagger\hat{A}_l\hat{W}_l\right]+
2\lambda\sum_{l=1}^{N-1}\left(1-{l\over N}\right)\hat{W}_{l+1}^\dagger \hat{W}_{l+1}$$
$$+\lambda (\hat{a}_1^\dagger +\hat{a}_L^\dagger )\left[
\sum_{l=1}^{N-1}\sqrt{1-{l\over N}}\hat{W}_{l}^\dagger\hat{W}_{l+1}\sqrt{C^1_{\hat{b}_1-\hat{\epsilon}}}+
\sum_{l=1}^{N}\sum_{k=1}^{N-1}\sqrt{1-{k\over N}}{\epsilon (k-l)\over |k-l|+1}\hat{W}_{l}^\dagger
\hat{A}_l \hat{A}_k^\dagger\hat{W}_{k+1} \right.
$$
$$\left. +\sum_{l=1}^{N-1}\sqrt{1-{l\over N}}\hat{W}_{l}^\dagger\hat{A}_l^\dagger\hat{A}_l\hat{W}_{l+1}\right].$$
Note that this Hamiltonian preserves both the number of open strings attached to the bound state
and the number of columns (= number of sphere giants). This is not the case at higher orders in ${1\over N}$ -
as discussed in section 3.1.2, there is a subleading term which allows the open string to occupy the first box
in the third column (the open string moves to occupy the first site in the giant lattice).
The effective field theory limit ($1\ll b_1 $) of this Hamiltonian is
$$ H_{boundary}=
\lambda (\hat{a}_1+\hat{a}_L)\left[
\sum_{l=1}^{N-1}\sqrt{1-{l\over N}}\hat{W}_{l+1}^\dagger\hat{W}_l+
\sum_{l=1,\, l\ne k}^{N}\sum_{k=1}^{N-1}\sqrt{1-{k\over N}}{1\over k-l}\hat{W}_{k+1}^\dagger
\hat{A}_k\hat{A}_l^\dagger \hat{W}_l \right]
$$
$$+
2\lambda\sum_{l=1}^{N-1}\left(1-{l\over N}\right)\hat{W}_{l+1}^\dagger \hat{W}_{l+1}
%
+\lambda (\hat{a}_1^\dagger +\hat{a}_L^\dagger )\left[
\sum_{l=1}^{N-1}\sqrt{1-{l\over N}}\hat{W}_{l}^\dagger\hat{W}_{l+1}\right.$$
\begin{equation}
\left. +
\sum_{l=1,\, l\ne k}^{N}\sum_{k=1}^{N-1}\sqrt{1-{k\over N}}{1\over k-l}\hat{W}_{l}^\dagger
\hat{A}_l \hat{A}_k^\dagger\hat{W}_{k+1} \right].
\label{last}
\end{equation}

The giant lattice we have been employing is not dynamical.
Further, at leading order, the number of sphere giants is conserved so that the Hamiltonians we have written are
rather simple.
In contrast to this, even at leading order, if we use this giant lattice to describe the dynamics of AdS giants
the number of particles on the giant lattice is not fixed. 
Using the dual lattice developed in appendix C, we find that at leading order the number of 
particles on the dual lattice is fixed and
the lattice is not dynamical. Thus, at leading order, the description of AdS gaints using the language of appendix C
seems to be the simplest. For this reason, we will not pursue the AdS giant dynamics using our present description.   
Finally, in the effective field theory limit, for $n$ sphere giants in the boundstate and a single
open string attached to the boundstate, it is simple to check that the dynamics is described by (\ref{last}). To obtain this result, we have
assumed that $n=O(1)$ and $n\ll b_i$, $i=0,1,...,n-1.$ See appendix B for further details on the effective field theory limit of boundstates
of three or four sphere or AdS giants.

\section{Toy Model}

To get some insight into the Hamiltonians we have obtained above, we will study a simple
toy model problem in this section: the case that the string has a single site, i.e. the
open string word only has 2 $Y$s in it. Toy models of this type were introduced in \cite{Berenstein:2006qk}
to study single excited sphere giants and used in \cite{adsgiant} to study single excited AdS giants.
Even with a single site, we are not able to solve the energy eigenvalue problem for a single excited sphere
or AdS giant analytically. For a single site, the numerical computations of the energy eigenvalues and
eigenkets is straight forward. One of the conclusions we reach, based on our numerical results, is
that backreaction on the membrane can not be neglected. The results\cite{Berenstein:2006qk},\cite{adsgiant} in 
the large $N$ limit, after neglecting back reaction, suggest a continuum of states separated from the ground state by a gap.
Since back reaction is neglected in these studies, the energy eigenvalue problem amounts to diagonalizing an
infinite dimensional matrix. In our approach, the number of $Z$s is finite so that there are only a finite number of
possible states for the giant/string system. Thus, our energy eigenvalue problem entails diagonalizing a finite
dimensional matrix. If in the boundary interaction terms
we hold the value of $\alpha =\sqrt{1-{b_0\over N}}$ (in the case of the sphere giant)
or $\alpha =\sqrt{1+{a_0\over N}}$ (in the case of the AdS giant)
fixed, we are neglecting the change in the giants momentum i.e. we are neglecting back reaction. In this case, 
even though we still diagonalize a finite dimensional matrix, we find good agreement with the results 
of\cite{Berenstein:2006qk},\cite{adsgiant}. Once back reaction is included, the gap disappears so that 
including back reaction seems to imply both a quantitative and a qualitative change of the result obtained ignoring 
back reaction. 

Although our results are suggestive on this point, things are not completely clear: indeed,
we compute the expectation value of the number operator for these energy eigenstates and
find that the planar approximation assumed when computing the open string word contractions is not accurate.
This implies that the single site results obtained from our Hamiltonian can not be trusted.

\subsection{Single Sphere Giant}

Our numerical analysis entails diagonalizing the matrix representation of the Hamiltonian. 
The system we consider has a total of $K$ $Z$s in the string/giant system. 
The ket with zeros everywhere except the $i$th entry
is the state with $K-i+1$ $Z$s on the giant and $J=i-1$ $Z$s on the string.
The matrix representation
of the hop off interaction is given by
$$ -\hat{A}^\dagger \hat{a}\sqrt{1-{K-\hat{J}\over N}}=\left[\matrix{
0 &\sqrt{1-{K-1\over N}}  &0                          &\dots   &0  &0\cr
0 &0                      &\sqrt{1-{K-2\over N}}      &\dots   &0  &0\cr
0 &0                      &0                          &\dots   &0  &0\cr
: &:                      &:                          &:       &:  &:\cr
0 &0                      &0                          &\dots   &0  &\sqrt{1-{K-K\over N}}\cr
0 &0                      &0                          &\dots   &0  &0 }\right].$$
This is a $K+1\times K+1$ matrix.
It is now straight forward to obtain the matrix representation of the Hamiltonian describing
a single excited sphere giant with a single string attached
$$H=\lambda\left[\matrix{
2\left( 1-{K-1\over N}\right) &2\sqrt{1-{K-1\over N}}         &0                             &\dots   &0  &0\cr
2\sqrt{1-{K-1\over N}}        &2+2\left( 1-{K-2\over N}\right)  &2\sqrt{1-{K-2\over N}}        &\dots   &0  &0\cr
0                             &2\sqrt{1-{K-2\over N}}         &2+2\left( 1-{K-3\over N}\right) &\dots   &0  &0\cr
:                             &:                              &:                             &:       &:  &:\cr
0                             &0                              &0                             &\dots   &4  &2\cr
0                             &0                              &0                             &\dots   &2  &2
}\right].$$
\begin{figure}[t]\label{fig:cgraph2}
\begin{center}
\includegraphics[height=6cm,width=9cm]{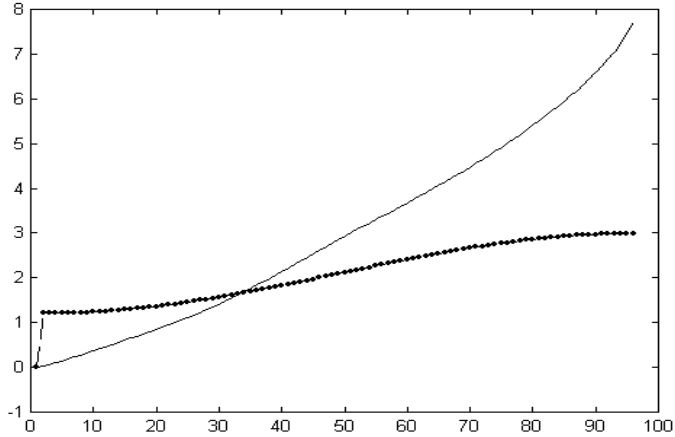}
\caption{The energy spectra for a single string attached to a single sphere giant. The plot shows $E_n$ versus $n$.
The energy is measured in units of $\lambda$. There are a total of 95 $Z$s in the string/brane
system and $N=100$. The solid curve shows the result obtained after backreaction is included. The result obtained
ignoring back reaction is plotted as a dashed line and the analytic
formula of \cite{Berenstein:2006qk} is plotted as a series of dots. The dashed curve is barely visible under the dots indicating
superb agreement between our numerical result and the result of \cite{Berenstein:2006qk}.} 
\end{center}
\end{figure}
One of the things we would like to establish, is the importance of back reaction. Towards this end, we have also constructed a
Hamiltonian that ignores the effects of back reaction. To ignore the effects of back reaction, we have kept the number of $Z$s on
the giant fixed, equal to $K$. Thus, ignoring back reaction, our hop off interaction, for example, is given by
$$ -\hat{A}^\dagger \hat{a}\sqrt{1-{K\over N}}=\left[\matrix{
0 &\sqrt{1-{K\over N}}  &0                          &\dots   &0  &0\cr
0 &0                      &\sqrt{1-{K\over N}}      &\dots   &0  &0\cr
0 &0                      &0                          &\dots   &0  &0\cr
: &:                      &:                          &:       &:  &:\cr
0 &0                      &0                          &\dots   &0  &\sqrt{1-{K\over N}}\cr
0 &0                      &0                          &\dots   &0  &0 }\right].$$
In the case that we ignore backreaction, we should be able to compare to the results of \cite{Berenstein:2006qk}. An important
difference between our work and that of \cite{Berenstein:2006qk}, is that in \cite{Berenstein:2006qk} the matrix representation
of the Hamiltonian is an infinite dimensional matrix. This is simply because an infinite number of $Z$s can hop off the giant and onto
the string. Our matrix representation for the Hamiltonian is a $K+1\times K+1$ matrix, corresponding to the fact that a maximum
of $K$ $Z$ fields can hop onto the string. Despite this difference, we find convincing agreement between our numerical results ignoring
back reaction and the analytic formula of \cite{Berenstein:2006qk}
$$ E(k)=2\lambda (1+2\alpha\cos (k)+\alpha^2),\qquad 0\le k\le \pi .$$
In figure 5 we have shown the energy spectra for $K=95$ and $N=100$. Our undeformed result is in perfect agreement with the analytic
result of \cite{Berenstein:2006qk}. Note that we are comparing normalizable states (the dots in figure 5) to states from the continuum
(the dashed line in figure 5). Since our system is described by a finite Hilbert space, our states are always normalizable. In support
of our assumption that this is a sensible comparison, note that the portion of the dashed curve describing continuum states is hidden
by the dots.
The result obtained when backreaction is taken into account is noticeably different from the result
obtained when back reaction is ignored. In particular, note that the mass gap obtained when backreaction is ignored, disappears when
back reaction is included.

\begin{figure}[t]\label{fig:cgraph2}
\begin{center}
\includegraphics[height=6cm,width=9cm]{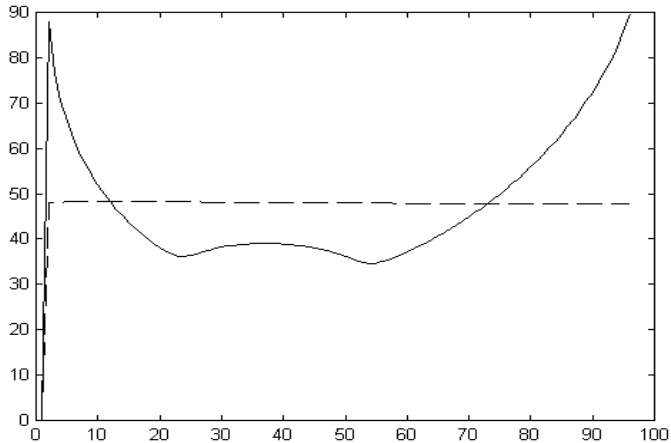}
\caption{The expectation value $\langle \hat{J}\rangle$ versus $n$, for a single string attached to a single sphere giant, is shown. 
There are a total of 95 $Z$s in the string/brane system and $N=100$. 
The solid curve shows the result obtained taking backreaction into account.
The result obtained ignoring backreaction is plotted as a dashed line.} 
\end{center}
\end{figure}
Our Hamiltonian is not exact. One approximation we have made is to assume that we need only sum planar diagrams when contracting the open 
string words. This amounts to assuming that the number of $Z$s on the open string ($=J$) is very much less than $\sqrt{N}$, so that
$J^2/N\ll 1$. We can easily compute $\langle \hat{J}\rangle$ numerically and see if the planar approximation is indeed accurate. In figure 6
we have plotted $\langle \hat{J}\rangle$.
Whether or not backreaction is included, $\langle \hat{J}\rangle$ is never much below 40 which is well outside the domain of validity of our Hamiltonian. 
Our Hamiltonian simply does not provide a valid description of the single site toy model, except for the ground state. 
Further, for these values of the parameters $K,N$, 
the interpretation of our system as an open string attached to a brane is not valid. 

An interesting question to ask is what is the time scale of the instability: Starting with a state corresponding to a string with a finite
number of $Z$s between the $Y$s, how long would it take before the dynamics is no longer captured by our Hamiltonian? If this time scale is 
long enough, one might be able to ignore non-planar effects for small time measurements\footnote{We thank David Berenstein for explaining this
to us.}. To estimate this time scale, recall the quantum brachistochrone problem: Given an initial
quantum state $|\psi_I\rangle$ and a final quantum state $|\psi_F\rangle$, how does one achieve the transfomation 
$|\psi_I\rangle\to |\psi_F\rangle =e^{-iHt/\hbar}|\psi_I\rangle$ in the shortest possible time? The optimization is 
with respect to the Hamiltonian, subject to the
constraint that the difference between smallest and largest eigenvalues of $H$ are held fixed. In Hermittian quantum mechanics, such a 
transformation always requires a non-zero amount of time. The optimal time is\cite{brach}
$$ t_{\rm instability}={2\over \Delta E}\arccos |\langle\psi_F|\psi_I\rangle | ,$$
where $\Delta E$ is the difference between the smallest and largest eigenvalues.
\begin{figure}[t]\label{fig:Ndep}
\begin{center}
\includegraphics[height=6cm,width=9cm]{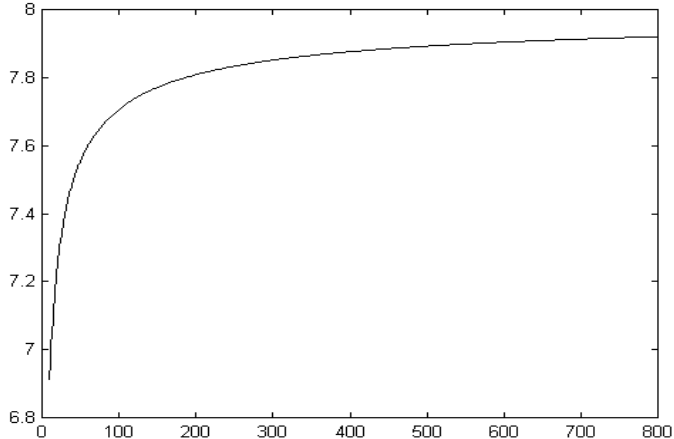}
\caption{The largest eigenvalue of the energy spectra for a single string attached to a single sphere giant, as a function of $N$.} 
\end{center}
\end{figure}

Figure 7 shows the numerical result for the largest eigenvalue as a function of $N$. From our numerical results, we read off
$\Delta E\approx 7.9 \lambda$. The final state has many $Z$'s; the initial state has very few $Z$'s. It is thus natural to approximate
$|\langle\psi_F|\psi_I\rangle |\approx 0$ and hence
$$ t_{\rm instability}={2\over \Delta E}\arccos |\langle\psi_F|\psi_I\rangle | 
\approx {2\over 7.9\times\lambda}{\pi\over 2}\approx {0.4\over\lambda} .$$
The interpretation of this result is straight forward: increasing $\lambda$ corresponds to increasing the string
tension. In this case, the string has a greater mass and thus offers increased resistance when the membrane tries to drag
it in non-geodesic motion. 

\subsection{Single AdS Giant}
\begin{figure}[t]\label{fig:cgraph2}
\begin{center}
\includegraphics[height=6cm,width=9cm]{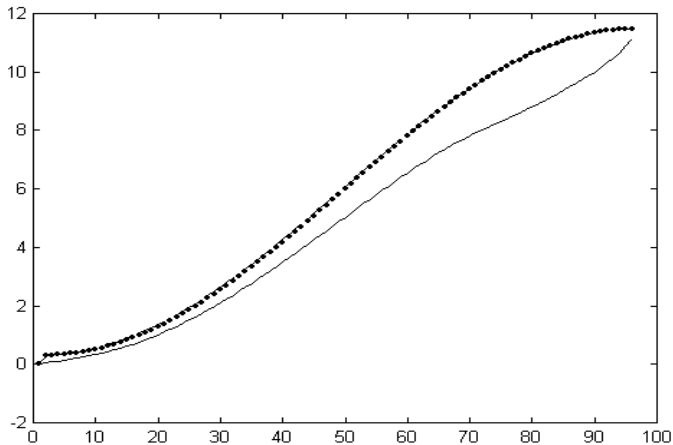}
\caption{The energy spectra for a single string attached to a single AdS giant. The plot shows $E_n$ versus $n$.
The energy is measured in units of $\lambda$. There are a total of 95 $Z$s in the string/brane
system and $N=100$. The solid curve shows the result obtained after backreaction is included. The result obtained
ignoring back reaction is plotted as a dashed line and the analytic
formula of \cite{adsgiant} is plotted as a series of dots. There is clearly
superb agreement between our numerical result and the result of \cite{adsgiant}.} 
\end{center}
\end{figure}

In this subsection we will determine the energy spectrum for a single string attached to an AdS giant, again by numerical
diagonalization of the matrix representation of the Hamiltonian. If the string plus AdS giant system has a total of $K$ $Z$
fields, the Hamiltonian is again a $K+1\times K+1$ matrix. The difference between the AdS and sphere giant problems is due
to the fact that the boundary interactions are different. For the AdS giant, the hop off interaction is given by
$$ \hat{A}^\dagger \hat{a}\sqrt{1+{K-\hat{J}\over N}}=\left[\matrix{
0 &\sqrt{1+{K-1\over N}}  &0                          &\dots   &0  &0\cr
0 &0                      &\sqrt{1+{K-2\over N}}      &\dots   &0  &0\cr
0 &0                      &0                          &\dots   &0  &0\cr
: &:                      &:                          &:       &:  &:\cr
0 &0                      &0                          &\dots   &0  &\sqrt{1+{K-K\over N}}\cr
0 &0                      &0                          &\dots   &0  &0 }\right].$$
It is now straightforward to determine the matrix representation of the Hamiltonian. We are again interested in determining the
importance of including the effects of backreaction on the AdS giant. To ignore the effects of back reaction, we again keep the
number of $Z$s on the brane fixed, leading to the hop off interaction
$$ \hat{A}^\dagger \hat{a}\sqrt{1+{K\over N}}=\left[\matrix{
0 &\sqrt{1+{K\over N}}  &0                          &\dots   &0  &0\cr
0 &0                      &\sqrt{1+{K\over N}}      &\dots   &0  &0\cr
0 &0                      &0                          &\dots   &0  &0\cr
: &:                      &:                          &:       &:  &:\cr
0 &0                      &0                          &\dots   &0  &\sqrt{1+{K\over N}}\cr
0 &0                      &0                          &\dots   &0  &0 }\right].$$
For the case that backreaction is ignored, we can compare to the results of \cite{adsgiant}. Just as for the case of an open
string attached to a sphere giant, an important
difference between our work and that of \cite{adsgiant}, is that in \cite{adsgiant} the matrix representation
of the Hamiltonian is an infinite dimensional matrix. Again, this is simply because an infinite number of $Z$s can hop off the giant and onto
the string. Our matrix representation for the Hamiltonian is a $K+1\times K+1$ matrix, corresponding to the fact that a maximum
of $K$ $Z$ fields can hop onto the string. The analytic result of \cite{adsgiant} for the spectrum is
$$ E(k)=2\lambda (1-2\alpha\cos (k)+\alpha^2),\qquad 0\le k\le \pi .$$
\begin{figure}[t]\label{fig:cgraph2}
\begin{center}
\includegraphics[height=6cm,width=9cm]{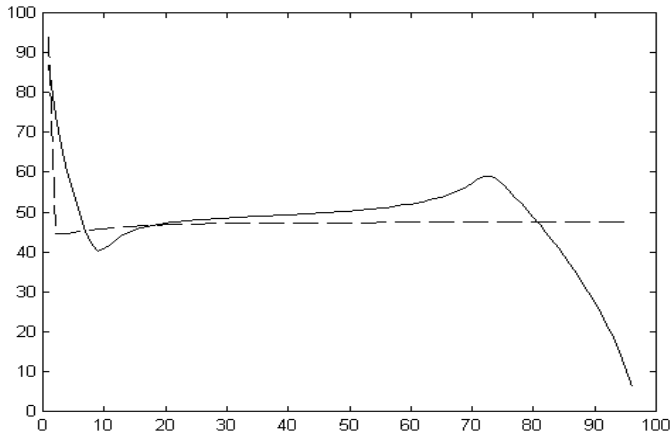}
\caption{The expectation value $\langle \hat{J}\rangle$ versus $n$, for a single string attached to a single AdS giant, is shown. 
There are a total of 95 $Z$s in the string/brane system and $N=100$. 
The solid curve shows the result obtained taking backreaction into account.
The result obtained ignoring backreaction is plotted as a dashed line.} 
\end{center}
\end{figure}

In figure 8 we have shown the spectra for $K=95$ and $N=100$. The agreement is again excellent\footnote{We thank Diego Correa for correcting an
error in $\alpha$ in the previous version of this paper.}.

Just as for the results we obtained for the excited sphere giants, the gap in the spectrum present when backreaction is ignored,
is removed when the effects of back reaction are included.

We can again check if our Hamiltonian is providing an accurate description of the physics. 
In figure 9 we have plotted $\langle \hat{J}\rangle$ versus $n$. 
It is clear that the planar approximation has broken down for all but the few highest energy states.
Again, for these values of the parameters $K,N$, the interpretation of our system as an open string attached to a brane 
is not valid. We are forced to conclude that our Hamiltonian does not provide an accurate description of a string with
a single site attached to an AdS giant. For the case of a single AdS giant, we obtain
$$ t_{\rm instability}={2\over \Delta E}\arccos |\langle\psi_F|\psi_I\rangle | 
\approx {2\over 11.6\times\lambda}{\pi\over 2}\approx {0.27\over\lambda} .$$

A few comments are in order. How are we to interpret the fact that our approximation breaks down? We have set up our description by
assuming that the operator we study is dual to a membrane with an open string attached. This implies that our operator can be 
decomposed into a ``membrane piece'' and a ``string piece''. These two pieces are treated very differently: when contracting the
membrane piece, all contractions are summed; when contracting the string piece, only planar contractions are summed. Contractions between
the two pieces are dropped. We have seen above, that a large number of $Z$s hop between the two $Y$s: our operator is simply not
dual to a state that looks like a membrane with an open string attached and our approximations are not valid. We are not
claiming that this operator does not have a planar limit - it should still be possible to study this operator using a systematic
$1/N$ expansion. Also, if one considered the same numerical study, but with $L\sim 10=\sqrt{N}$ $Y$s we would expect our Hamiltonian
to provide a suitable description. This problem appears to be too numerically expensive to perform in practice.

\section{The Semiclassical Limit}

In the previous section we have argued that our Hamiltonian does not accurately describe the dynamics of a single string attached to
a giant graviton, when the string has a single site. In this section we will consider the opposite limit in which we take $L\to\infty$.
This limit has been considered, for a single sphere giant in AdS$_5\times$S$^5$, in \cite{CuntzChain},\cite{Berenstein:2006qk}, for a
single sphere giant in a $\gamma$ deformed background in \cite{deMelloKoch:2005jg} and for a single AdS giant in AdS$_5\times$S$^5$ in \cite{adsgiant}.
In the limit, the dynamics of the Cuntz chain is governed by a semiclassical sigma model. This semiclassical sigma model coincides
with the Polyakov action describing an open string attached to the 
giant \cite{CuntzChain},\cite{deMelloKoch:2005jg},\cite{Berenstein:2006qk},\cite{adsgiant}.
This strongly suggests that the Cuntz chain Hamiltonian is relevant for the description of this $L\to\infty$ limit. In this section we
will study the semiclassical sigma models arising from the semiclassical limit of our Cuntz chain Hamiltonians.

To warm up, we will consider the case of a single sphere giant or a single AdS giant. 
We will employ the description developed in subsection 3.1.1 as this is, by far, the simplest description.
For the semiclassical limit, we take $L\to\infty$ and $\lambda\to\infty$ holding $\lambda/L^2$ fixed and small.
In addition, we put each site of the lattice into a coherent state of a Cuntz oscillator
$$|z\rangle =\sqrt{1-|z|^2}\sum_{n=0}^\infty z^n |n\rangle ,\qquad |z|<1.$$
The parameter for the coherent state of the $l$th lattice site is $z_l =r_l (t)e^{i\phi_l (t)}$.
In this article, we also allow the brane to be dynamical.
To obtain a semiclassical limit, we also put the brane in a coherent state, with parameter $Z=R(t)e^{i\Phi (t)}$.
The resulting action is given by\cite{Zhang:1990fy}
$$S=\int dt\left(
i\langle z_1,...,z_L;Z|{d\over dt}|z_1,...,z_L;Z\rangle - \langle z_1,...,z_L;Z|H|z_1,...,z_L;Z\rangle
\right).$$
The first term in the action and the bulk terms in the Cuntz chain Hamiltonian are the same for any brane
system that the open string is attached to and hence may be read from the results of\cite{CuntzChain}. 
The first term in the action becomes
$$i\langle z_1,...,z_L;Z|{d\over dt}|z_1,...,z_L;Z\rangle 
= -\sum_{l=1}^L {r_l^2\dot{\phi}_l\over 1-r_l^2}-{R^2\dot{\Phi}\over 1-R^2}$$
$$=-L\int_0^1 {\dot{\phi}(\sigma )r^2(\sigma )\over 1-r^2(\sigma)}d\sigma -{R^2\dot{\Phi}\over 1-R^2}.$$
The bulk terms of the Cuntz chain are
$$-\langle z_1,...,z_L;Z|\left[
2\lambda\sum_{l=1}^L a_l^\dagger a_l -\lambda\sum_{l=1}^{L-1}(a_l^\dagger a_{l+1}+a_la^\dagger_{l+1})
\right]|z_1,...,z_L;Z\rangle $$
$$= -2\lambda\sum_{l=1}^L \bar{z}_l z_l +\lambda\sum_{l=1}^{L-1}(\bar{z}_l z_{l+1}+z_l \bar{z}_{l+1})$$
$$=-L{\lambda\over L^2}\int_0^1\left[\left({\partial r\over\partial\sigma}\right)^2+r^2\left(
{\partial\phi\over\partial\sigma}\right)^2\right]d\sigma -\lambda\left[ r^2(1)+r^2(0)\right].$$
To obtain the above integral representations, we have made use of the Euler-Maclaurin formula. There are corrections to
the integrals we have written above, expressed in terms of derivatives of the function evaluated at the endpoints. These corrections
will need to be taken into account when the first ${1\over L}$ corrections are computed. 

The remaining boundary interactions in the sigma model are dependent on the details of the specific brane system we
study. In the next two subsections we will consider the interactions relevant for a single AdS or a single sphere giant.
In the third subsection, we argue that the AdS giant is unstable. Finally, in the last subsection we consider the semiclassical
limit of a boundstate of giants.

A final comment is in order. Since our Hamiltonian preserves the number of $Z$ fields in the giant plus string system (denoted by $K$)
we will look for solutions that minimize the energy and have a sharp classical value for $K$. Concretely, we do this by setting the coherent
state expectation value of $\hat{K}$ equal to $K$.

\subsection{Single Sphere Giant}

The coherent state expectation value of the boundary interaction Hamiltonian given in section 3.4, gives
the following contribution to the action
$$-2\lambda Z\bar{Z}\left[1-{K\over N}+{1\over N}\sum_{l=1}^L {\bar{z}_l z_l\over 1-\bar{z}_l z_l}\right]
-\lambda\left[\bar{Z}(z_1+z_L)+Z(\bar{z}_1+\bar{z}_L)\right]
\sqrt{1-{K\over N}+{1\over N}\sum_{l=1}^L {\bar{z}_l z_l\over 1-\bar{z}_l z_l}}$$
$$=-2\lambda R^2\left[ 1-{K\over N}+{L\over N}\int_0^1 {r^2 (\sigma)\over 1-r^2(\sigma )}d\sigma\right]$$
$$-\lambda\left[ \bar{Z}(z(0)+z(1))+Z(\bar{z}(0)+\bar{z}(1))\right]
 \sqrt{1-{K\over N}+{L\over N}\int_0^1 {r^2 (\sigma)\over 1-r^2(\sigma )}d\sigma} .$$
This result is not exact. The number operator $\hat{b}_0$ appears in the Hamiltonian; we have replaced it
by its coherent state expectation value
$$\langle\hat{b}_0\rangle ={K\over N}-{1\over N}\sum_{l=1}^L {\bar{z}_l z_l\over 1-\bar{z}_l z_l}.$$
The semi-classical sigma model action describing a single string attached to a sphere giant
graviton is
$$ S=\int L_\sigma dt $$
$$ L_\sigma =-L\int_0^1 {\dot{\phi}r^2\over 1-r^2}d\sigma -{R^2\dot{\Phi}\over 1-R^2}
-L{\lambda\over L^2}\int_0^1\left[\left({\partial r\over\partial\sigma}\right)^2+r^2\left(
{\partial\phi\over\partial\sigma}\right)^2\right]d\sigma -\lambda\left[ r^2(1)+r^2(0)\right]$$
$$-\lambda\left[ \bar{Z}(z(0)+z(1))+Z(\bar{z}(0)+\bar{z}(1))\right]
\sqrt{1-{K\over N}+{L\over N}\int_0^1 {r^2\over 1-r^2}d\sigma} $$
$$-2\lambda R^2\left[ 1-{K\over N}+{L\over N}\int_0^1 {r^2\over 1-r^2}d\sigma\right].$$
In the above action, $Z$ and $z_l$ are not independent - they are coupled by the constraint
$$K=\sum_{n=1}^\infty (\hat{A}^\dagger)^n \hat{A}^n +\sum_{l=1}^L \sum_{n=1}^\infty (\hat{a}_l^\dagger)^n \hat{a}_l^n $$
which says that the total number of $Z$s is equal to the number of $Z$s on the giant plus the number of $Z$s
on the string. The coherent state expectation value of the constraint is
$$ K={\bar{Z}Z\over 1-\bar{Z}Z}+L\int_0^1 {r^2\over 1-r^2}d\sigma ={\bar{Z}Z\over 1-\bar{Z}Z}+J ,$$
where we have introduced the coherent state expectation value of the number of $Z$s on the string, $J\equiv\langle\hat{J}\rangle$.
This is easily solved to eliminate $|Z|$
$$\bar{Z}Z=R^2=1-{1\over K+1-L\int_0^1 {r^2\over 1-r^2}d\sigma }.$$
Using this constraint, we obtain 
$$ L_\sigma =-L\int_0^1 {\dot{\phi}r^2\over 1-r^2}d\sigma -{R^2\dot{\Phi}\over 1-R^2}
-L{\lambda\over L^2}\int_0^1\left[\left({\partial r\over\partial\sigma}\right)^2+r^2\left(
{\partial\phi\over\partial\sigma}\right)^2\right]d\sigma -\lambda\left[ r^2(1)+r^2(0)\right]$$
$$-2\lambda\left[ r(0)\cos (\phi (0)-\Phi)+r(1)\cos (\phi (1)-\Phi)\right]\sqrt{1-{K\over N}+{J\over N}}
\sqrt{K-J\over 1+K-J}$$
$$-{2\lambda\over 1+K-J} \left[ K-J-{(K-J)^2\over N}\right],$$
If we now shift $\phi (\sigma )\to \phi(\sigma )+\Phi $, the giant and string dynamics decouple so that we finally obtain
a sigma model expressed only in terms of $r(\sigma )$ and $\phi (\sigma )$\footnote{We have dropped the term
$-{R^2\dot{\Phi}\over 1-R^2}$ from the Lagrangian, as it is not needed to obtain the string dynamics.}
$$ L_\sigma =-L\int_0^1 {\dot{\phi}r^2\over 1-r^2}d\sigma 
-L{\lambda\over L^2}\int_0^1\left[\left({\partial r\over\partial\sigma}\right)^2+r^2\left(
{\partial\phi\over\partial\sigma}\right)^2\right]d\sigma -\lambda\left[ r^2(1)+r^2(0)\right]$$
$$-2\lambda\left[ r(0)\cos (\phi (0))+r(1)\cos (\phi (1))\right]\sqrt{1-{K\over N}+{J\over N}}
\sqrt{K-J\over 1+K-J}$$
\begin{equation}
-{2\lambda\over 1+K-J} \left[ K-J-{(K-J)^2\over N}\right].
\label{finalmodel}
\end{equation}
In the limit we study $K=O(N)$, $K\gg J$ and $\alpha =\sqrt{1-{K\over N}}=O(1)$, our system can be interpreted as a string attached to brane
and further, we expect that back reaction will be a subleading effect. $K$ is a fixed parameter which we may therefore choose to be $O(N)$.
$J$ is determined by the dynamics, and thus the issue of how large it is compared to $K$ is a dynamical question.
It is natural to expect that $K\gg J$, since the energy contribution from the boundary terms is minimized for small values of $J$.
In this limit (\ref{finalmodel}) becomes
$$ L_\sigma =-L\int_0^1 {\dot{\phi}r^2\over 1-r^2}d\sigma 
-L{\lambda\over L^2}\int_0^1\left[\left({\partial r\over\partial\sigma}\right)^2+r^2\left(
{\partial\phi\over\partial\sigma}\right)^2\right]d\sigma -\lambda\left[ r^2(1)+r^2(0)\right]$$
$$-2\lambda\left[ r(0)\cos (\phi (0))+r(1)\cos (\phi (1))\right]\alpha
 - 2\lambda \alpha^2 ,$$
which is in perfect agreement with the sigma model action obtained in \cite{CuntzChain}.
This action can be obtained directly as a limit of the Polyakov action in a certain gauge, as
demonstrated in \cite{CuntzChain}. This suggests that our Hamiltonian does provide a reliable 
description of this semiclassical limit. 

The corrections to $L_\sigma$ due to back reaction are $O({J\over K})$. We know that, at most 
we can tolerate $J\sim\sqrt{N}$ - beyond this our description breaks down. To correct it we would
have to go beyond the planar approximation employed when contracting the open string words.
Further, our giant has $K=O(N)$. Thus, when our description is valid
$O({J\over K})=O({1\over L})$, so that the ${1\over L}$ corrections to our sigma model action are the same
size as the corrections due to back reaction and the corrections coming from the Euler-Maclaurin formula. 
All of these ${1\over L}$ corrections need to be included when the effects of back reaction are studied.

\subsection{Single AdS Giant}

For a single AdS giant, the coherent state expectation value of the boundary interaction Hamiltonian given
in section 3.4, gives the following contribution to the action
$$-2\lambda Z\bar{Z}\left[1+{K\over N}-{1\over N}\sum_{l=1}^L {\bar{z}_l z_l\over 1-\bar{z}_l z_l}\right]
+\lambda\left[\bar{Z}(z_1+z_L)+Z(\bar{z}_1+\bar{z}_L)\right]
\sqrt{1+{K\over N}-{1\over N}\sum_{l=1}^L {\bar{z}_l z_l\over 1-\bar{z}_l z_l}}$$
$$=-2\lambda R^2\left[ 1+{K\over N}-{J\over N}\right]
+\lambda\left[ \bar{Z}(z(0)+z(1))+Z(\bar{z}(0)+\bar{z}(1))\right]
 \sqrt{1+{K\over N}-{J\over N}} .$$
This is again not an exact result - we have replaced $\hat{a}_0$ by its coherent state expectation value.
Thus, the semi-classical sigma model action describing a single string attached to an AdS giant
graviton is
$$ S=\int L_\sigma dt $$
$$ L_\sigma =-L\int_0^1 {\dot{\phi}(\sigma )r^2(\sigma )\over 1-r^2(\sigma)}d\sigma -{R^2\dot{\Phi}\over 1-R^2}
-L{\lambda\over L^2}\int_0^1\left[\left({\partial r\over\partial\sigma}\right)^2+r^2\left(
{\partial\phi\over\partial\sigma}\right)^2\right]d\sigma -\lambda\left[ r^2(1)+r^2(0)\right]$$
$$+\lambda\left[ \bar{Z}(z(0)+z(1))+Z(\bar{z}(0)+\bar{z}(1))\right]
\sqrt{1+{K\over N}-{J\over N}}
-2\lambda R^2\left[ 1+{K\over N}-{J\over N}\right].$$
In the above action, $Z$ and $z_l$ are again not independent - they are coupled by the same constraint that we
obtained for the sphere giant, and hence we may again set
$$\bar{Z}Z=R^2=1-{1\over K+1-J}.$$
After employing the constraint to eliminate $R$, and shifting $\phi(\sigma )\to\phi(\sigma )+\Phi$ which again decouples
the string and the brane dynamics, we obtain
$$ L_\sigma =-L\int_0^1 {\dot{\phi}r^2\over 1-r^2}d\sigma 
-L{\lambda\over L^2}\int_0^1\left[\left({\partial r\over\partial\sigma}\right)^2+r^2\left(
{\partial\phi\over\partial\sigma}\right)^2\right]d\sigma -\lambda\left[ r^2(1)+r^2(0)\right]$$
$$+2\lambda\left[ r(0)\cos (\phi (0))+r(1)\cos (\phi (1))\right]\sqrt{1+{K\over N}-{J\over N}}
\sqrt{K-J\over 1+K-J}$$
\begin{equation}
-{2\lambda\over 1+K-J} \left[ K-J+{(K-J)^2\over N}\right].
\label{finalAdSmodel}
\end{equation}
We only expect this Lagrangian to be an accurate description of the dynamics
in the limit that $K=O(N)$, $K\gg J$ and $\alpha =\sqrt{1+{K\over N}}=O(1)$, which 
is the limit we are considering.
We will see below that this is not a valid assumption.
As discussed above for the sphere giant, the size of $J$ is a dynamical question. In contrast to what we
found for the sphere giant, the boundary terms in this Lagrangian are minimized for large values of $J$. 
Continuing anyway with the above assumption, our system can be interpreted as a string attached to a brane
and further, back reaction will be a subleading effect. In this limit (\ref{finalAdSmodel}) becomes
$$ L_\sigma =-L\int_0^1 {\dot{\phi}r^2\over 1-r^2}d\sigma 
-L{\lambda\over L^2}\int_0^1\left[\left({\partial r\over\partial\sigma}\right)^2+r^2\left(
{\partial\phi\over\partial\sigma}\right)^2\right]d\sigma -\lambda\left[ r^2(1)+r^2(0)\right]$$
$$+2\lambda\left[ r(0)\cos (\phi (0))+r(1)\cos (\phi (1))\right]\alpha - 2\lambda\alpha^2 . $$
If we now shift $\phi(\sigma )\to\phi(\sigma) +\pi$ then this action becomes identical in form to the action describing the 
single string attached to a sphere giant. Of course, one very important difference is that here $\alpha\ge 1$; for the
string attached to a sphere giant, $\alpha\le 1$.

\subsection{Interpretation of the Single Giant Results}

In this section we will study solutions to the sigma models of sections 5.1 and 5.2, which correspond to point like strings
for which $\dot{r}=r'=0$ and $\dot{\phi}=\phi'=0$. The bulk equations of motion (which are the same for the two types of
giants)
$$ {\lambda\over L}r''={L\dot{\phi}r\over (1-r^2)^2}+{\lambda r(\phi')^2\over L},$$
$$ {r\dot{r}\over (1-r^2)^2}+\partial_\sigma\left({\lambda\over L^2}r^2 \phi'\right)=0,$$
are clearly satisfied. The boundary conditions for the sphere and the AdS giants are different. 
Consider the case of the sphere giant first.
The boundary terms in (\ref{finalmodel}) are minimized if we apply the boundary conditions
\begin{equation}
\phi (0)=\phi (1)=\pi,\qquad r(0)=r(1)= \sqrt{1-{K\over N}+{J\over N}}
\sqrt{K-J\over 1+K-J}. 
\label{sphereboundary}
\end{equation}
If we ignore back reaction (set $J=0$ in the above equation) and ${1\over K}$ corrections, we find
\begin{equation} 
r(0)=r(1)=\sqrt{1-{K\over N}},
\label{sphereb}
\end{equation}
with $K$ now equal to the momentum of the giant. For the AdS giant, the boundary terms in (\ref{finalAdSmodel}) vanish
if we require
\begin{equation}
\phi(0)=\phi(1)=0,\qquad r(0)=r(1)=\sqrt{1+{K\over N}-{J\over N}}
\sqrt{K-J\over 1+K-J}.
\label{adsboundary}
\end{equation}
Ignoring back reaction we find
\begin{equation}
r(0)=r(1)=\sqrt{1+{K\over N}},
\label{adsb}
\end{equation}
with $K$ now equal to the momentum of the giant.

To interpret these boundary conditions, recall how the AdS$_5\times$S$^5$ solution is recovered from the LLM description.
The AdS$_5\times$S$^5$ geometry corresponds to a circular droplet boundary condition on the $y=0$ plane, parameterized
by $(x_1,x_2)$ (see section 2.3 of \cite{Lin:2004nb}). Introduce radial coordinates $(r,\phi)$ on this plane. 
The $r$ and $y$ coordinates are related to $\rho$ (the radial variable of AdS$_5$ in global coordinates) and
$\theta$ (one of the angles of the S$^5$) by $y=r_0\sinh\rho\sin\theta$ and $r=r_0\cosh\rho\cos\theta$, where 
$r_0=R_{{\rm AdS}_5}^2=R_{{\rm S}^5}^2$.
The sphere giants are located at $\rho=0$ and $\cos\theta =\sqrt{1-{K\over N}}$ so that $y=0$ and $r=\sqrt{1-{K\over N}}$. 
The AdS giants are located at $\theta=0$ and $\cosh\rho = \sqrt{1+{K\over N}}$ so that $y=0$ and $r=\sqrt{1+{K\over N}}$. 
This matches beautifully with  (\ref{sphereb}) and (\ref{adsb}).
We thus obtain a clear geometrical interpretation of our coherent state parameter $z=re^{i\phi}$ - the $r$ in our coherent
state parameter is the radial direction on the $y=0$ LLM plane. With this identification, our strings are localized 
on the $y=0$ plane which is colored black or white. The sphere giant sits in a black region; the AdS giant in a white region. 
In a black region, the $S^3$ in AdS$_5$ has shrunk to zero size; in a white region, the $S^3$ in the $S^5$ has shrunk to zero. This
implies that in a white region, (for our AdS giant) we can't have a string with angular momentum 
on the $S^3$ contained in the $S^5$. If this interpretation is correct, our description (\ref{finalAdSmodel}) must fail. 

In \cite{Berenstein:2006qk} a potential source of a D-brane instability was discovered. The giant graviton couples to the background
RR flux $F_5$. This coupling produces a Lorentz force acting on the brane and consequently, the giant does not undergo free motion.
The string, which does not couple to $F_5$ and hence would undergo geodesic motion, thus feels a force from the brane as the brane
drags it along. If this force is enough to overcome the tension of the string, the string will be stretched to large lengths, allowing
smaller loops to pinch off. In this way, the brane would decay into gravitational radiation. We conjecture that the AdS giants are
unstable against this decay, which is the source of the failure of our description (\ref{finalAdSmodel}).
\begin{figure}[t]\label{fig:cgraph2}
\begin{center}
\includegraphics[height=6cm,width=9cm]{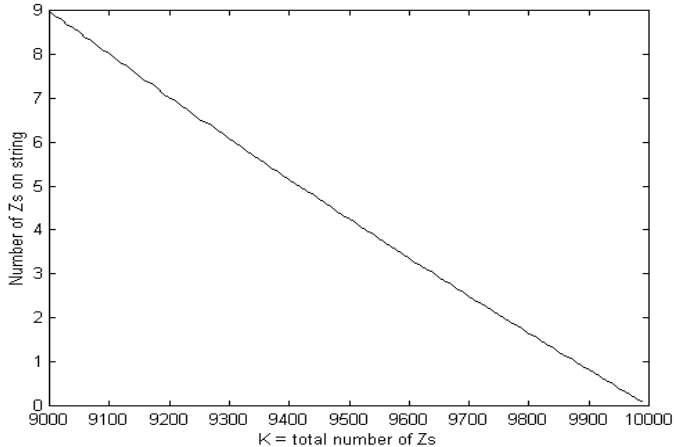}
\caption{The expectation value $\langle \hat{J}\rangle$ versus $K$, for a single string attached to a single sphere giant, is shown. 
There are a total of K $Z$s in the string/brane system, the string has $L=80$ sites and $N=10000$.} 
\end{center}
\end{figure}

To provide some evidence for this interpretation, return to (\ref{sphereboundary}) and (\ref{adsboundary}). Our point string ansatz for
the sphere giant is
$$ r=a,\qquad \phi=\pi .$$
With this ansatz,
\begin{equation} 
J={La^2\over 1-a^2}.
\label{numop}
\end{equation}
It is now possible to solve (\ref{sphereboundary}) and (\ref{numop}) simultaneously to determine $J$.
If back reaction effects are negligible, we expect that $J\ll\sqrt{N}$. From figure 10, it is clear that back reaction
is indeed negligible.

For the case of the AdS giant, we determine $J$ by solving (\ref{adsboundary}) and (\ref{numop}) simultaneously.
If the AdS giant is unstable, we would expect the effects of back reaction to be large, and hence $J$ should
be large. Of course, this implies that the dynamics is no longer described by our Hamiltonian. From figure 11 it is clear that the
AdS giant suffers from significant back reaction, supporting our conjecture that the AdS giants are unstable.
We are not able to verify this conjecture by a detailed study of this instability; this is outside the validity of our
description which fails as soon as ${J^2\over N}\sim 1$. 
Besides the fact that $J$ is so large that our sigma model description can not be trusted, we see that $J>K$ which 
indicates that this is not a valid solution. 
\begin{figure}[t]\label{fig:cgraph2}
\begin{center}
\includegraphics[height=6cm,width=9cm]{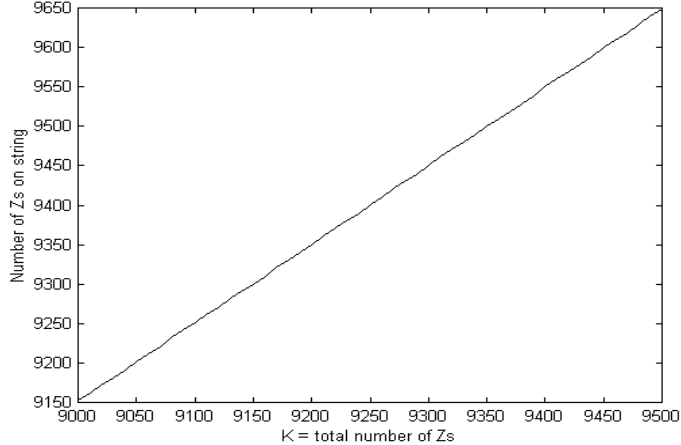}
\caption{The expectation value $\langle \hat{J}\rangle$ versus $K$, for a single string attached to a single AdS giant, is shown. 
There are a total of K $Z$s in the string/brane system, the string has $L=80$ sites and $N=10000$.} 
\end{center}
\end{figure}

\subsection{Bound State of Sphere Giants}

In this section we consider an open string attached to a bound state of two sphere giants.
In the case that a single open string attaches to a bound state of giant gravitons, the Gauss
law forces both endpoints of the string to attach to the same brane. To stretch strings between
two giants, we need at least two open strings attached to the bound state. In the case that open
strings stretch between the giants, we would expect to see a force between the branes. For a
single string attached to the bound state, we should be able to verify that we can recover the
physics of a single string attached to a single brane, when the two branes are well separated.

Our strategy is again to consider point like string solutions $\dot{r}=r'=0$, $\dot{\phi}=\phi'=0$
to the sigma model. As in previous sections, the bulk equations of motion are clearly satisfied,
so that we need only focus on the boundary terms in the Hamiltonian. The wave function $|\Psi\rangle$
for this excited bound state will be a direct product of a ket describing the open string with a ket
describing the giant. If we denote the point string state by $|z=re^{i\phi}\rangle_{\rm string}$,
we can write 
$$ |\Psi\rangle =|z=re^{i\phi}\rangle_{\rm string}\otimes 
                 \sum_{l,p=1}^{N-1} c_{lp}\hat{W}^\dagger_l \hat{A}_p^\dagger |0\rangle ,$$
where the constants $c_{lp}$ need to be determined. To simplify our analysis, we will assume that
$b_1\gg 1$. With this assumption, the boundary Hamiltonian\footnote{We have added $2\lambda z\bar{z}$
- which comes from the (string) bulk Hamiltonian (\ref{bulk}) - to this boundary Hamiltonian.}, 
when acting on the state $|\Psi\rangle$, is
$$ H_{boundary}=
2\lambda z\left[
\sum_{l=1}^{N-1}\sqrt{1-{l\over N}}\hat{W}_{l+1}^\dagger\hat{W}_l+
\sum_{l=1,\, l\ne k}^{N}\sum_{k=1}^{N-1}\sqrt{1-{k\over N}}{1\over k-l}\hat{W}_{k+1}^\dagger
\hat{A}_k\hat{A}_l^\dagger \hat{W}_l \right]
$$
$$+2\lambda\sum_{l=1}^{N-1}\left(1-{l\over N}\right)\hat{W}_{l+1}^\dagger \hat{W}_{l+1}
+2\lambda z\bar{z} $$
$$+2\lambda \bar{z}\left[
\sum_{l=1}^{N-1}\sqrt{1-{l\over N}}\hat{W}_{l}^\dagger\hat{W}_{l+1}+
\sum_{l=1,\, l\ne k}^{N}\sum_{k=1}^{N-1}\sqrt{1-{k\over N}}{1\over k-l}\hat{W}_{l}^\dagger
\hat{A}_l \hat{A}_k^\dagger\hat{W}_{k+1} \right].
$$

For the case of a single string attached to a giant, we put the giant into a coherent state and fixed the coherent state parameter
so that the boundary contribution to the energy vanished. In this section we will show that essentially the same approach works for
a boundstate of two giants. Using our giant lattice notation, a coherent state for a single sphere giant, can be written as
$$ |Z\rangle =\sum_l Z^l\hat{W}_l^\dagger |0\rangle .$$
Motivated by this observation, we have studied the state
$$ |\Psi\rangle = \sum_{l,p=1}^{N-1} c_{lp}\hat{W}^\dagger_l \hat{A}_p^\dagger |0\rangle ={\cal N}
\sum_{l_1=1}^{N-1}\sum_{l_2=1}^{N-1}(Z_1)^{l_1}(Z_2)^{l_2}\hat{W}^\dagger_{l_1}\hat{A}^\dagger_{l_2}|0\rangle ,$$
where
%
%
$$ {\cal N}^{-2}={Z_1\bar{Z}_1-(Z_1\bar{Z}_1)^{N}\over 1-Z_1\bar{Z}_1}{Z_2\bar{Z}_2-(Z_2\bar{Z}_2)^{N}\over 1-Z_2\bar{Z}_2} .$$
%
%
To compute the coherent state expectation value of the terms in the Hamiltonian that do not depend on the $\hat{A}_p$
oscillator
$$ H_{W}=
2\lambda z
\sum_{l=1}^{N-1}\sqrt{1-{l\over N}}\hat{W}_{l+1}^\dagger\hat{W}_l
+ 2\lambda\sum_{l=1}^{N-1}\left(1-{l\over N}\right)\hat{W}_{l+1}^\dagger \hat{W}_{l+1}
+2\lambda z\bar{z}$$
$$+2\lambda \bar{z} 
\sum_{l=1}^{N-1}\sqrt{1-{l\over N}}\hat{W}_{l}^\dagger\hat{W}_{l+1},
$$
we will replace $l$ by its coherent state expectation value (this is the same approximation employed in sections 5.1 and 5.2)
$$\langle\hat{l}\rangle ={1\over 1-Z_1 \bar{Z}_1}.$$
For a giant graviton, we are interested in the case that $N(1-Z_1 \bar{Z}_1)$ is $O(1)$ so that $Z\bar{Z}_1=1-\alpha N^{-1}+O(N^{-2}).$
This sets the radius of our giant graviton
$$ R^2= R_{S^5}^2{1\over N(1-Z_1 \bar{Z}_1)}=R_{S^5}^2{1\over \alpha }.$$
It is now straight forward to verify that, to leading order at large $N$, we have
\begin{equation}
\langle\Psi |H_W |\Psi\rangle = 2\lambda (z\bar{Z}_1+ \bar{z}Z_1)\sqrt{1\over N(1-Z_1 \bar{Z}_1)} 
+2\lambda Z_1\bar{Z}_1\left[{1\over N(1-Z_1 \bar{Z}_1)}\right] +2\lambda z\bar{z}.
\label{Benergy}
\end{equation}
We can set (\ref{Benergy}) to zero by choosing
\begin{equation} 
Z_1=R_1 e^{i\Phi_1},\quad \Phi_1=\phi+\pi,
\label{twosolved}
\end{equation}
$$ r=R_1 \sqrt{1-{1\over N(1-R_1^2)}}=\sqrt{1-{1\over N(1-R_1^2)}},$$
with the last equality holding at leading order in $N$. This is a very natural result: the string is located precisely on 
the radius of the orbit of the giant in spacetime. Next, consider
$$ \langle H_4\rangle =\lambda \langle\Psi |\sum_{l=1,\, l\ne k}^{N}\sum_{k=1}^{N-1}\sqrt{1-{k\over N}}{1\over k-l}\hat{W}_{k+1}^\dagger
\hat{A}_k\hat{A}_l^\dagger \hat{W}_l |\Psi\rangle ,$$
$$ \approx\lambda \sqrt{1-{1\over N (1-Z_2\bar{Z}_2)}}\sum_{l=1,\, l\ne k}^{N}\sum_{k=1}^{N-1}{1\over k-l}\langle\Psi | \hat{W}_{k+1}^\dagger
\hat{A}_k\hat{A}_l^\dagger \hat{W}_l |\Psi\rangle ,$$
$$ = {\cal N}^2 \lambda \bar{Z}_1\sqrt{1-{1\over N (1-Z_2\bar{Z}_2)}}
\sum_{k,l=1,\, k\ne l}^{N-1} (\bar{Z}_1 Z_2)^k (\bar{Z}_2 Z_1)^l {1\over k-l}.$$
If we choose $Z_2=R_2e^{i\Phi_1}$, the above expectation value vanishes. It is now easy to see that, with this choice for $Z_2$ and the
choice (\ref{twosolved}) we have
$$ \langle\Psi |H_{boundary} |\Psi\rangle = 0,$$
so that the contribution to the energy coming from $H_{boundary}$ is minimized.

The only loose end is to fix the sum of the number of $Z$s in the giant boundstate plus the number of $Z$s in the string to $K$.
After taking the coherent state expectation value of the constraint
$$ K=\sum_{l=1}^{N-1}l\hat{W}_l^\dagger\hat{W}_l+\sum_{k=1}^{N-1}k\hat{A}_k^\dagger\hat{A}_k +\hat{J},$$
we obtain (recall that $K$ and $L$ are given quantum numbers of the operator - they are not determined by the dynamics)
$$K={1\over 1-Z_1\bar{Z}_1}+{1\over 1-Z_2\bar{Z}_2}+{Lr^2\over 1-r^2}.$$
This is a single equation for the two parameters $R_2$ and $r$, indicating that our solution has a single free parameter. This is
expected - it specifies how we share the momentum between the two giants in the boundstate.

\section{Conclusions}

In this article, we have given methods that determine the Cuntz chain Hamiltonians describing the dynamics
of open strings attached to giant gravitons. These Hamiltonians are accurate to first order in $g_{YM}^2$.
The bulk term of the Hamiltonian has been obtained previously. The contribution of the present article is
to obtain an explicit expression for the boundary interactions. There are boundary interactions which allow
the string and the membrane to exchange momentum. We have managed to obtain an explicit expression for the
back reaction on the membrane as a result of these interactions. Although the interactions are rather 
complicated, we have found a natural interpretation for the coefficients which appear. For example, there are
coefficients that are responsible
for gracefully switching off certain interactions as the branes become coincident. Further, we have found an
``effective field theory limit" in which the Hamiltonians simplify considerably.

The operators we consider are labeled by Young diagrams; open strings are denoted by filling the boxes of
the Young diagram with the label of the open string. We have only considered attaching a single string to each system of
giants in this article. One interesting feature of our results, is that the Young diagrams labeling the operators
have a clear geometrical interpretation. Indeed, one of the processes allowed by the boundary interactions involves
a string detaching from the brane to which it is attached and reattaching to a second brane in the system. In terms
of the labels for the operators, the open string hops from one box in the Young diagram, to a different box, and
in the process it changes both the row and the column it is in. We have
found clear signals that we should interpret the number of boxes separating the box that the string starts in from the
box that the string lands up in, as we move on the right hand side of the Young diagram, as a distance. 
This distance is related to the radial coordinate of the two dimensional $y=0$ plane on which the LLM boundary
conditions are specified\cite{Lin:2004nb},\cite{Balasubramanian:2005mg}.
The interaction
displays an inverse dependence on this distance. The effective theory describing these open strings should be a 
Yang-Mills theory, local not on the space on which the original field theory is defined, but rather on the $3+1$
dimensional worldvolume
of the brane we are describing\cite{Balasubramanian:2004nb}. This new space should emerge from the matrix degrees of freedom of the ${\cal N}=4$
super Yang-Mills theory. The ${1\over r}$ potential which would arise from the exchange of massless particles in three
spatial dimensions, thus looks rather natural.

The Polyakov action for both closed\cite{Kr} and open strings\cite{CuntzChain} emerge as a semi-classical limit from a spin chain 
(or equivalently for us, from a Cuntz chain)
that can be derived directly from the gauge theory. In this work we have managed to provide a complete account of the back reaction 
of the string on the membrane. In particular, we have introduced a Cuntz oscillator chain (which is equivalent to a spin chain) for 
the giant graviton itself. This Cuntz oscillator chain keeps track of the ``motion of the corners'' of the Young diagram, which 
describes the giant bound state. It is natural to expect that the semi-classical limit of this Cuntz chain will make contact with
membrane dynamics in the dual gravitational description.

We have tried to build a toy model in which we consider a string with a single site. 
The advantage of the toy model is that it is numerically tractable.
For this ``short string" toy model the planar approximation used in computing the contractions between open string words is not valid. 
Thus, our Hamiltonians do not accurately describe this ``short string limit" and the numerical results are not to be trusted.

Finally, we have considered the opposite limit in which the number of sites in the open string $L$ is taken to infinity 
$L\sim O(\sqrt{N})$. In this semi classical limit, the dynamics of the Cuntz chain is governed by a sigma model.
We have argued that the description of strings attached to sphere giants is reliably captured by the sigma model dynamics.
This was argued by showing that back reaction on the giant is a small effect. In contrast to this, the back reaction on 
an AdS giant is so large that the use of the sigma model to describe the open string dynamics is not valid. Based on this
result, we conjecture that the AdS giant is unstable against gravitational decay, although a detailed study of this question
is not within reach of our sigma model description.
Finally, we studied an open string attached to a bound state of two sphere giants.
In the case that a single open string attaches to a bound state of giant gravitons, the Gauss
law forces both endpoints of the string to attach to the same brane. We have recovered the
physics of a single string attached to a single brane, when the two branes are well separated,
as expected.

There are a number of directions in which our results can be extended. 
It would be interesting to extend our results
to include the case that two strings are attached to the system of giants. When we have two (or more)
strings attached to the giant graviton bound state, we can have strings
stretching between different branes. Studying this system would allow us to compute the
force between two branes. 

We have initiated a study of the dynamics of our Cuntz chain Hamiltonians. A natural question to
ask is if this dynamics is integrable or not? Following the discussion of \cite{Berenstein:2006qk},
if this is the case, integrability might not be realized by a Bethe Ansatz. See \cite{Sam} for a recent
discussion of this question.

Recently an extremely interesting proposal for determining the metric of LLM geometries from closed string sigma models
constructed as the semiclassical limit of Cuntz chain dynamics was given in \cite{SamMetric}. Can this be extended to open
string dynamics for open strings attached to giants probing the general ${1\over 2}$ BPS (LLM) geometry?

Given the technology we have developed for operators dual to excited giant gravitons, it may be possible to construct
the string field theory describing open strings attached to giant gravitons, systematically. A powerful
framework that has already given impressive results for closed string field theory\cite{friends} exploits the methods of collective field
theory\cite{coll}.

Finally, we have restricted ourselves to the $SU(2)$ sector in this article. This corresponds to studying strings
with two angular momenta on the sphere. One could generalize our analysis to the full $SO(6)$ excitations on the sphere
and further, to include spin in the AdS space. One could also consider attaching strings to giants that preserve less
supersymmetry\cite{aristos}.

{\vskip 0.4truecm}

\noindent
{\it Acknowledgements:} 
We would like to thank Michael Abbott, 
Simon Connell, Sera Cremonini, Aristomenis Donos, Antal Jevicki, Charles Kasl, 
Jeff Murugan, Sanjaye Ramgoolam, Joao Rodrigues and especially David Berenstein, 
for pleasant discussions and/or helpful correspondence. 
This work is based upon research supported by the South African Research Chairs
Initiative of the Department of Science and Technology and National Reseach Foundation.
Any opinion, findings and conclusions or recommendations expressed in this material
are those of the authors and therefore the NRF and DST do not accept any liability
with regard thereto.

\vfill\eject

\appendix

\section{An Identity}

In this appendix, we derive an identity that can be used to obtain the Cuntz chain Hamiltonian that accounts for the
$O(g_{YM}^2)$ correction (coming from the $F$ terms) to the anomalous dimension of our operators.
Our starting point is the restricted Schur polynomial
$$\chi^{(1)}_{R,R'}(Z,W)={1\over (n-1)!}\sum_{\sigma\in S_n}\Tr_{R'}\left(\Gamma_{R}\big[\sigma\big]\right)
Z^{i_1}_{i_{\sigma(1)}}Z^{i_2}_{i_{\sigma(2)}}\cdots Z^{i_{n-1}}_{i_{\sigma(n-1)}}W^{i_n}_{i_{\sigma(n)}}.$$
$R$ is a representation of $S_n$; $\Gamma_{R}\big[\sigma\big]$ is the matrix representing $\sigma$ in representation $R$.
Rewrite the above sum as a sum over the $S_{n-1}$ subgroup that leaves $n$ unchanged ($\sigma(n)=n$), and its cosets. After
rearranging the resulting expression a little, we obtain
\begin{eqnarray}
\nonumber
\chi^{(1)}_{R,R'}(Z,W)-\chi_{R'}(Z)\Tr (W) &=&
{1\over (n-1)!}\sum_{\sigma\in S_{n-1}}\left[
\Tr_{R'}\left(\Gamma_{R}\left[\sigma (n,1)\right]\right)
(ZW)^{i_1}_{i_{\sigma(1)}}Z^{i_2}_{i_{\sigma(2)}}\cdots Z^{i_{n-1}}_{i_{\sigma(n-1)}}\right.\\
\nonumber
& &+
\Tr_{R'}\left(\Gamma_{R}\left[\sigma (n,2)\right]\right)
Z^{i_1}_{i_{\sigma(1)}}(ZW)^{i_2}_{i_{\sigma(2)}}\cdots Z^{i_{n-1}}_{i_{\sigma(n-1)}}+....+\\
\label{terms}
& &+\left.
\Tr_{R'}\left(\Gamma_{R}\left[\sigma (n,n-1)\right]\right)
Z^{i_1}_{i_{\sigma(1)}}Z^{i_2}_{i_{\sigma(2)}}\cdots (ZW)^{i_{n-1}}_{i_{\sigma(n-1)}}\right] .
\end{eqnarray}
where $\chi_{R'}(Z)$ is the Schur polynomial. To obtain this result, use
$$ \chi_{R,R'}(Z) = {1\over (n-1)!}
   \sum_{\sigma\in S_{n-1}}\Tr_{R'}\left[\Gamma_{R}(\sigma)\right]
   Z^{i_1}_{i_{\sigma(1)}}Z^{i_2}_{i_{\sigma(2)}}\cdots Z^{i_{n-1}}_{i_{\sigma(n-1)}}$$
$$= {1\over (n-1)!}
   \sum_{\sigma\in S_{n-1}}\chi_{R'}(\sigma)
   Z^{i_1}_{i_{\sigma(1)}}Z^{i_2}_{i_{\sigma(2)}}\cdots Z^{i_{n-1}}_{i_{\sigma(n-1)}}$$
$$\equiv\chi_{R'}(Z).$$
Introduce the notation $W^+ = ZW$. Concentrate on the first term in (\ref{terms}). This term
can be rewritten as a sum over the $S_{n-2}$ subgroup of $S_{n-1}$, which comprises of the
permutations which leave $1$ fixed ($\sigma (1)=1$), and its cosets
\begin{eqnarray}
\nonumber
{1\over (n-1)!}& &\sum_{\sigma\in S_{n-1}}
\Tr_{R'}\left(\Gamma_{R}\left[\sigma (n,1)\right]\right)
(W^+)^{i_1}_{i_{\sigma(1)}}Z^{i_2}_{i_{\sigma(2)}}\cdots Z^{i_{n-1}}_{i_{\sigma(n-1)}}\\
\nonumber
&=& {1\over (n-1)!}\sum_{\sigma\in S_{n-2}}
\Tr_{R'}\left(\Gamma_{R}\left[\sigma (n,1)\right]\right)
\Tr(W^+)Z^{i_2}_{i_{\sigma(2)}}\cdots Z^{i_{n-1}}_{i_{\sigma(n-1)}}\\
\nonumber
& &+{1\over (n-1)!}\sum_{\sigma\in S_{n-2}}
\Tr_{R'}\left(\Gamma_{R}\left[\sigma (1,2)(n,1)\right]\right)
(W^+ Z)^{i_2}_{i_{\sigma(2)}}\cdots Z^{i_{n-1}}_{i_{\sigma(n-1)}}\\
\nonumber
& &+\cdots + {1\over (n-1)!}\sum_{\sigma\in S_{n-2}}
\Tr_{R'}\left(\Gamma_{R}\left[\sigma (1,n-1)(n,1)\right]\right)
Z^{i_2}_{i_{\sigma(2)}}\cdots (W^+ Z)^{i_{n-1}}_{i_{\sigma(n-1)}}
\end{eqnarray}
We can break $R'=\oplus_\alpha R''_{\alpha}$ where the 
sum runs over all representations $R''_\alpha$
that can be obtained from $R'$ by removing a single box.
The subgroup we sum over leaves both $n$ and $1$ inert so that
$$\Gamma_R\big[\tau\big]\Gamma_R\big[(n,1)\big]=\Gamma_R\big[(n,1)\big]\Gamma_R\big[\tau\big] .$$
By Schur's Lemma this implies that $\Gamma_R\big[(n,1)\big]$ is proportional to the identity
when acting on the $R''_\alpha$ subspace
\begin{equation}
\label{keyfact}
\langle a, R''_\alpha |\Gamma_R\big[(n,1)\big]|b,R_\alpha''\rangle =\lambda_\alpha \delta_{ab}.
\end{equation}
Decomposing the trace over $R'$ we have
$$ \Tr_{R'}\left(\Gamma_R\big[\tau (1,n-1)(n,1)\big]\right)
=\sum_{\alpha}\Tr_{R''_\alpha}\left(\Gamma_R\big[\tau (1,n-1)(n,1)\big]\right) .$$
Now, thanks to the block diagonal structure of $\Gamma_R \big[\tau\big]$ we know
$$\langle a,R''_\alpha |\Gamma_R \big[\tau\big]|b,R''_\beta\rangle\propto \delta_{\alpha\beta},$$
which allows us to write
\begin{eqnarray}
\nonumber
\sum_{\alpha}\Tr_{R''_\alpha}\left(\Gamma_R\big[\tau (1,n-1)(n,1)\big]\right)&=&
\sum_\alpha\sum_a \langle a,R''_\alpha |\Gamma_R\big[\tau (1,n-1)(n,1)\big]|a, R''_\alpha\rangle\\
\nonumber
&=& \sum_\alpha\sum_a \sum_b \langle a,R''_\alpha |\Gamma_R\big[\tau\big]|b,R''_\alpha\rangle
\langle b,R''_\alpha |\Gamma_R\big[(1,n-1)(n,1)\big]|a, R''_\alpha\rangle .
\end{eqnarray}
Now, lets introduce the notation
$$ P_{R\to R'\to R_\alpha''}|_{n,1}=\sum_b |b,R_\alpha''\rangle\langle b,R_\alpha''|.$$
Recall that $R$ is a representation of $S_n$, $R'$ is a representation of $S_{n-1}$ and $R_\alpha''$ is a representation of
$S_{n-2}$. The notation $|_{n,1}$ tells us how to make the projection: the $S_{n-1}$ subgroup is obtained from $S_n$ by taking
the subgroup of elements that leave $n$ fixed; $S_{n-2}$ is obtained from $S_{n-1}$ by taking the subgroup of elements that
leave 1 fixed. Using this new notation, it is clear that
$$\Gamma_R\big[(1,n-1)\big] P_{R\to R'\to R_\alpha''}|_{n,1}\Gamma_R\big[(1,n-1)\big]=P_{R\to R'\to R_\alpha''}|_{n,n-1},$$
so that
$$ P_{R\to R'\to R_\alpha''}|_{n,1}\Gamma_R\big[(1,n-1)\big]=\Gamma_R\big[(1,n-1)\big]P_{R\to R'\to R_\alpha''}|_{n,n-1}.$$
We now find
\begin{eqnarray}
\nonumber
& &\sum_\alpha\sum_a \sum_b \langle a,R''_\alpha |\Gamma_R\big[\tau\big]|b,R''_\alpha\rangle
\langle b,R''_\alpha | \Gamma_R\big[(1,n-1)(n,1)\big]|a, R''_\alpha\rangle\\
&=&\sum_\alpha\sum_a \langle a,R''_\alpha |\Gamma_R\big[\tau\big] P_{R\to R'\to R_\alpha''}|_{n,1}
\Gamma_R\big[(1,n-1)(n,1)\big]|a, R''_\alpha\rangle \\
\nonumber
&=& \sum_\alpha\sum_a \langle a,R''_\alpha |\Gamma_R\big[\tau\big]
\Gamma_R\big[(1,n-1)\big]P_{R\to R'\to R_\alpha''}|_{n,n-1}
\Gamma_R\big[(n,1)\big]|a, R''_\alpha\rangle \\
\nonumber
&=& \sum_\alpha\sum_a \langle a,R''_\alpha |\Gamma_R\big[\tau\big]
\Gamma_R\big[(1,n-1)\big]P_{R\to R'\to R_\alpha''}|_{n,n-1}
\left(\sum_\beta P_{R\to R'\to R_\beta''}|_{n,1}\right)
\Gamma_R\big[(n,1)\big]|a, R''_\alpha\rangle ,
\end{eqnarray}
where we have used that fact that $\sum_\beta P_{R\to R'\to R_\beta''}|_{n,1}$ acts as the identity on the $R'$ subspace.
Now, using (\ref{keyfact}) we obtain
\begin{eqnarray}
\nonumber
& &\sum_\alpha\sum_a \langle a,R''_\alpha |\Gamma_R\big[\tau\big]
\Gamma_R\big[(1,n-1)\big]P_{R\to R'\to R_\alpha''}|_{n,n-1}
\left(\sum_\beta P_{R\to R'\to R_\beta''}|_{n,1}\right)
\Gamma_R\big[(n,1)\big]|a, R''_\alpha\rangle\\
&=& \sum_\alpha\sum_a \lambda_\alpha \langle a,R''_\alpha |
\Gamma_R\big[\tau\big] \Gamma_R\big[(1,n-1)\big]
P_{R\to R'\to R_\alpha''}|_{n,n-1} |a, R''_\alpha\rangle\\
\nonumber
&=& \sum_\alpha\sum_a \lambda_\alpha \langle a,R''_\alpha |\Gamma_R\big[\tau\big]
P_{R\to R'\to R_\alpha''}|_{n,1} \Gamma_R\big[(1,n-1)\big] |a, R''_\alpha\rangle\\
\nonumber
&=& \sum_\alpha\sum_a \lambda_\alpha \Tr_{R''_\alpha}(\Gamma_R\big[\tau (1,n-1) \big]).
\end{eqnarray}
Thus, our final result is
$$ \Tr_{R'}\left(\Gamma_{R}\left[\tau (1,n-1)(n,1)\right]\right)=
\sum_\alpha\lambda_\alpha \Tr_{R''_\alpha}(\Gamma_R\big[\tau (1,n-1) \big]). $$

If we now consider $\Tr_{R'}\left(\Gamma_{R}\left[\tau (1,i)(n,1)\right]\right)$, and if
we restrict to the subgroup $S_{n-2}$ obtained by taking all the elements of $S_{n-1}$ that hold $i$ fixed, we find
$$ \Tr_{R'}\left(\Gamma_{R}\left[\tau (1,i)(n,1)\right]\right)=
\sum_\alpha\lambda_\alpha \Tr_{R''_\alpha}(\Gamma_R\big[\tau (1,i) \big]). $$
Thus, we now have
\begin{eqnarray}
\nonumber
{1\over (n-1)!}& &\sum_{\sigma\in S_{n-1}}
\Tr_{R'}\left(\Gamma_{R}\left[\sigma (n,1)\right]\right)
(W^+)^{i_1}_{i_{\sigma(1)}}Z^{i_2}_{i_{\sigma(2)}}\cdots Z^{i_{n-1}}_{i_{\sigma(n-1)}}\\
\nonumber
&=& {1\over (n-1)!}\sum_{\alpha}\lambda_\alpha\sum_{\sigma\in S_{n-2}}
\Tr_{R''_\alpha}\left(\Gamma_{R}\left[\sigma\right]\right)
\Tr(W^+)Z^{i_2}_{i_{\sigma(2)}}\cdots Z^{i_{n-1}}_{i_{\sigma(n-1)}}\\
\nonumber
& &+{1\over (n-1)!}\sum_{\alpha}\lambda_\alpha\sum_{\sigma\in S_{n-2}}
\Tr_{R''_\alpha}\left(\Gamma_{R}\left[\sigma (1,2)\right]\right)
(W^+ Z)^{i_2}_{i_{\sigma(2)}}\cdots Z^{i_{n-1}}_{i_{\sigma(n-1)}}\\
\nonumber
& &+\cdots + {1\over (n-1)!}\sum_{\alpha}\lambda_\alpha\sum_{\sigma\in S_{n-2}}
\Tr_{R''_\alpha}\left(\Gamma_{R}\left[\sigma (1,n-1)\right]\right)
Z^{i_2}_{i_{\sigma(2)}}\cdots (W^+ Z)^{i_{n-1}}_{i_{\sigma(n-1)}}\\
\nonumber
&=&{1\over (n-1)!}\sum_{\alpha}\lambda_\alpha\sum_{\sigma\in S_{n-1}}
\Tr_{R''_\alpha}\left(\Gamma_{R'}\left[\sigma \right]\right)
(W^+)^{i_1}_{i_{\sigma(1)}}Z^{i_2}_{i_{\sigma(2)}}\cdots Z^{i_{n-1}}_{i_{\sigma(n-1)}}\\
\nonumber
&=&{1\over n-1}\sum_\alpha\lambda_\alpha\chi^{(1)}_{R',R''_\alpha}(Z,W^+).
\end{eqnarray}
It is not difficult to see that each of the $n-1$ terms on the right hand side
of (\ref{terms}) makes exactly the same contribution, so that 
$$ \chi^{(1)}_{R,R'}(Z,W)-\chi_{R'}(Z)\Tr (W) =
\sum_\alpha\lambda_\alpha\chi^{(1)}_{R',R''_\alpha}(Z,W^+) .$$
All that remains is to compute $\lambda_\alpha$. This was done in \cite{jelena}.
The result is (the subgroup of which $R_\alpha''$ in the next formula is a representation
is obtained by holding $n$ and then $n-1$ fixed)
$$\lambda_\alpha ={1\over d_{R_\alpha''}}\Tr_{R''_\alpha}(\Gamma_R\left[ (n,n-1)\right])=
{1\over c_{RR'}-c_{R'R''_\alpha}},$$
where $c_{RR'}$ is the weight of the box that must be removed from $R$ to obtain $R'$ and
$c_{R'R''_\alpha}$ is the weight of the box that must be removed from $R'$ to obtain $R''_\alpha$.
The formula that we will make use of, is
\begin{equation}
\label{goodone}
\chi^{(1)}_{R,R'}(Z,W)-\chi_{R'}(Z)\Tr (W) =
\sum_\alpha{1\over c_{RR'}-c_{R'R''_\alpha}}\chi^{(1)}_{R',R''_\alpha}(Z,ZW) .
\end{equation}
With a little thought, the motivated reader can convince herself that a very similar argument can
be constructed to show that
$$\chi^{(1)}_{R,R'}(Z,W)-\chi_{R'}(Z)\Tr (W) =
\sum_\alpha{1\over c_{RR'}-c_{R'R''_\alpha}}\chi^{(1)}_{R',R''_\alpha}(Z,WZ) .$$

To conclude this section, we will illustrate this formula with a few examples.
We will employ the graphical notation introduced in \cite{jelena}. 
Then, for example, (\ref{goodone}) says
$$\chi_{\young({\,}{\,}{\,}{w},{\,}{\,})}-\chi_{\young({\,}{\,}{\,},{\,}{\,})}\Tr (w)
=\chi_{\young({\,}{\,}{x},{\,}{\,})}+{1\over 3}\chi_{\young({\,}{\,}{\,},{\,}{x})},\qquad x=w^+$$
$$\chi_{\young({\,}{\,},{\,}{\,},{\,},{w})}-\chi_{\young({\,}{\,},{\,}{\,},{\,})}\Tr (w)
=-\chi_{\young({\,}{\,},{\,}{\,},{x})}-{1\over 3}\chi_{\young({\,}{\,},{\,}{x},{\,})},\qquad x=w^+$$

\subsection{Numerical Test}

The main result of this article is the formula (\ref{goodone}). Indeed, this formula determines the hop 
off interaction. The hop on interaction then follows from the hop off interaction by Hermitian conjugation 
and the kissing interaction by composing the hop on and the hop off interactions. Thus, the complete boundary
interaction and the corresponding back reaction on the brane are determined by (\ref{goodone}). Given the 
importance of this formula, we have tested it numerically. In this subsection we will explain
the check we have performed.

The formula (\ref{goodone}) is an identity between restricted Schur polynomials. It must be true if we evaluate
it for {\it any}\footnote{In particular, not necessarily Hermitian.} numerical value of the matrices $Z$ and $W$. 
Our check entails evaluating both sides of (\ref{goodone}), for a number of different matrices $W$ and $Z$, to 
check the equality. Evaluating a restricted Schur polynomial entails evaluating:

$$ $$

\noindent
(i) The restricted character $\Tr_{R'}\left(\Gamma_{R}\big[\sigma\big]\right)$: This was done by explicitely 
constructing the matrices $\Gamma_{R}\big[\sigma\big]$. Each representation used was obtained by induction. 
One induces a reducible representation; the irreducible representation that participates was isolated using
projection operators built from the Casimir obtained by summing over all two cycles. See appendix B.2 of
\cite{jelena} for more details. The resulting irreducible representations were tested by verifying the 
multiplication table of $S_n$.

\noindent
(ii) The trace $\Tr (\sigma Z^{\otimes n-1} W)=Z^{i_1}_{i_{\sigma(1)}}Z^{i_2}_{i_{\sigma(2)}}\cdots 
Z^{i_{n-1}}_{i_{\sigma(n-1)}} W^{i_n}_{i_{\sigma(n)}}$: for any given $\sigma\in S_n$ this trace is 
easily expressed as a product of traces of powers of $Z$ and $W$.

$$ $$

Our code verified (\ref{goodone}) for all possible restricted Schur polynomials that could be constructed with 
$R\in S_5$, $R'\in S_4$ and $R''\in S_3$. This gives 12 independent tests in total.

\section{Boundstates of Giant Gravitons}

In this appendix we will give the results for the hop off interaction, for boundstates of three or four giant gravitons. 
Our motivation for doing this is to exhibit a general structure, in the effective field theory limit, that can be used to 
write down the hop off interaction for an arbitrary number of boundstates. 
Further, we would like to argue that the features discussed in section 3.1.2 for the
boundstate of two giants hold for the general giant boundstate. Given the hop off interaction,
one can construct the full Hamiltonian exactly as we have done in section 3. In this appendix, we use the following notation:
$$ W^{(1)}=W(\{n_1,n_2,...,n_L\}), $$
$$ W^{(2)}=W(\{n_1-1,n_2,...,n_L\})\quad {\rm or} W^{(2)}=W(\{n_1,n_2,...,n_L-1\}).$$
The two possibilities for $W^{(2)}$ above correspond to the freedom to hop off either end point of the string.

\subsection{Boundstate of three sphere giants}

The hop off interaction for a boundstate of three sphere giants is
$$ H|\{n_{b_0}=1, n_{b_0+b_1}=1,n_{b_0+b_1+b_2}=\bar{1}\};W^{(1)} \rangle=$$
$$ 
-\lambda\left[ \sqrt{1-{b_0+b_1+b_2\over N}}  A_s
|\{n_{b_0}=1,n_{b_0+b_1}=1, n_{b_0+b_1+b_2+1}=\bar{1}\};W^{(2)} \rangle \right.$$
$$ + \sqrt{1-{b_0+b_1-1\over N}}
B_s |\{n_{b_0}=1,n_{b_0+b_1+1}=\bar{1}, n_{b_0+b_1+b_2}=1\};W^{(2)} \rangle$$
$$ \left.  +  \sqrt{1-{b_0-2\over N}}C_s
|\{n_{b_0+1}=\bar{1},n_{b_0+b_1}=1, n_{b_0+b_1+b_2}=1\};W^{(2)} \rangle \right],$$
$$ $$
$$ $$
$$ H|\{n_{b_0}=1,n_{b_0+b_1}=\bar{1},n_{b_0+b_1+b_2}=1\};W^{(1)} \rangle=$$
$$ 
-\lambda\left[ \sqrt{1-{b_0+b_1+b_2\over N}}  D_s
|\{n_{b_0}=1,n_{b_0+b_1}=1, n_{b_0+b_1+b_2+1}=\bar{1}\};W^{(2)} \rangle \right.$$
$$ + \sqrt{1-{b_0+b_1-1\over N}}
E_s |\{n_{b_0}=1,n_{b_0+b_1+1}=\bar{1}, n_{b_0+b_1+b_2}=1\};W^{(2)} \rangle$$
$$ \left.  +  \sqrt{1-{b_0-2\over N}}F_s
|\{n_{b_0+1}=\bar{1},n_{b_0+b_1}=1, n_{b_0+b_1+b_2}=1\};W^{(2)} \rangle \right],$$
$$ $$
$$ $$
$$ H|\{n_{b_0}=\bar{1},n_{b_0+b_1}=1,n_{b_0+b_1+b_2}=1\};W^{(1)} \rangle$$
$$ 
-\lambda\left[ \sqrt{1-{b_0+b_1+b_2\over N}}  G_s
|\{n_{b_0}=1,n_{b_0+b_1}=1, n_{b_0+b_1+b_2+1}=\bar{1}\};W^{(2)} \rangle \right.$$
$$ + \sqrt{1-{b_0+b_1-1\over N}}
H_s |\{n_{b_0}=1,n_{b_0+b_1+1}=\bar{1}, n_{b_0+b_1+b_2}=1\};W^{(2)} \rangle$$
$$ \left.  +  \sqrt{1-{b_0-2\over N}}I_s
|\{n_{b_0+1}=\bar{1},n_{b_0+b_1}=1, n_{b_0+b_1+b_2}=1\};W^{(2)} \rangle \right],$$
where
$$ A_s=-{\sqrt{b_1+b_2+1}\sqrt{b_1+b_2+3}\sqrt{b_2}\sqrt{b_2+2}\over (b_2+1)(b_1+b_2+2)},$$
$$ B_s=\sqrt{b_1+2\over b_1+1}\sqrt{b_1+b_2+1\over b_1+b_2+2}{1\over b_2+1},$$
$$ C_s=\sqrt{b_2\over b_2+1}\sqrt{b_1\over b_1+1}{1\over b_1+b_2+2},$$
$$ D_s=-\sqrt{b_1\over b_1+1}\sqrt{b_1+b_2+3\over b_1+b_2+2}{1\over b_2+1},$$
$$ E_s=-{\sqrt{b_1}\sqrt{b_1+2}\sqrt{b_2}\sqrt{b_2+2}\over (b_2+1)(b_1+1)},$$
$$ F_s=\sqrt{b_2+2\over b_2+1}\sqrt{b_1+b_2+1\over b_1+b_2+2}{1\over b_1+1},$$
$$ G_s=-\sqrt{b_1+2\over b_1+1}\sqrt{b_2+2\over b_2+1}{1\over b_1+b_2+2},$$
$$ H_s=-\sqrt{b_2\over b_2+1}\sqrt{b_1+b_2+3\over b_1+b_2+2}{1\over b_1+1},$$
$$ I_s-{\sqrt{b_1+b_2+1}\sqrt{b_1+b_2+3}\sqrt{b_1}\sqrt{b_1+2}\over (b_1+1)(b_1+b_2+2)} .$$ 
These expressions look ugly. However, things simplify dramatically in the effective field theory limit.
For example, $A_s$ very rapidly approaches 1 as either $b_1$ or $b_2$ is increased. To illustrate this
point, we have plotted $A_s$ as a function of $b_2$.
\begin{figure}[t]\label{fig:cgraph2}
\begin{center}
\includegraphics[height=6cm,width=9cm]{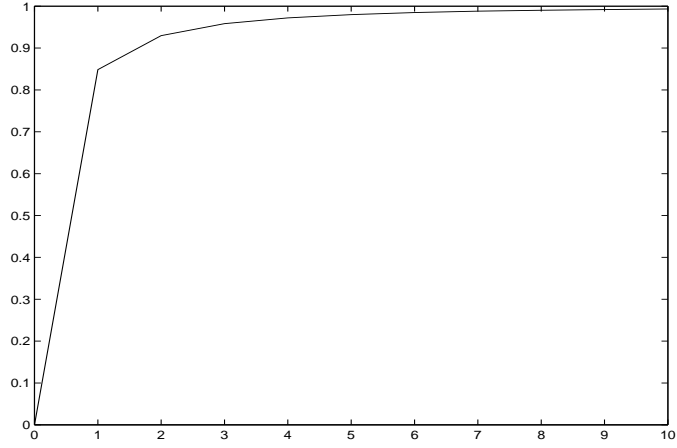}
\caption{A plot of $A_s$ versus $b_2$ for $b_1=2$. It is clear that $A_s$ very rapidly approaches 1
as $b_2$ is increased.} 
\end{center}
\end{figure}

In the effective field theory limit, the hop off interaction is well approximated by
$$ H|\{n_{b_0}=1, n_{b_0+b_1}=1,n_{b_0+b_1+b_2}=\bar{1}\};W^{(1)} \rangle=$$
$$ 
-\lambda\left[-\sqrt{1-{b_0+b_1+b_2\over N}}
|\{n_{b_0}=1,n_{b_0+b_1}=1, n_{b_0+b_1+b_2+1}=\bar{1}\};W^{(2)} \rangle \right.$$
$$ + \sqrt{1-{b_0+b_1 \over N}}{1\over b_2}
|\{n_{b_0}=1,n_{b_0+b_1+1}=\bar{1}, n_{b_0+b_1+b_2}=1\};W^{(2)} \rangle$$
$$ \left.  +  \sqrt{1-{b_0\over N}}{1\over b_1+b_2}
|\{n_{b_0+1}=\bar{1},n_{b_0+b_1}=1, n_{b_0+b_1+b_2}=1\};W^{(2)} \rangle \right],$$
$$ $$
$$ $$
$$ H|\{n_{b_0}=1,n_{b_0+b_1}=\bar{1},n_{b_0+b_1+b_2}=1\};W^{(1)} \rangle=$$
$$ 
-\lambda\left[-\sqrt{1-{b_0+b_1+b_2\over N}}  {1\over b_2}
|\{n_{b_0}=1,n_{b_0+b_1}=1, n_{b_0+b_1+b_2+1}=\bar{1}\};W^{(2)} \rangle \right.$$
$$-\sqrt{1-{b_0+b_1 \over N}}
|\{n_{b_0}=1,n_{b_0+b_1+1}=\bar{1}, n_{b_0+b_1+b_2}=1\};W^{(2)} \rangle$$
$$ \left.  +  \sqrt{1-{b_0\over N}}{1\over b_1}
|\{n_{b_0+1}=\bar{1},n_{b_0+b_1}=1, n_{b_0+b_1+b_2}=1\};W^{(2)} \rangle \right],$$
$$ $$
$$ $$
$$ H|\{n_{b_0}=\bar{1},n_{b_0+b_1}=1,n_{b_0+b_1+b_2}=1\};W^{(1)} \rangle$$
$$ 
\lambda\left[ \sqrt{1-{b_0+b_1+b_2\over N}}  {1\over b_1+b_2}
|\{n_{b_0}=1,n_{b_0+b_1}=1, n_{b_0+b_1+b_2+1}=\bar{1}\};W^{(2)} \rangle \right.$$
$$ + \sqrt{1-{b_0+b_1 \over N}}{1\over b_1}
|\{n_{b_0}=1,n_{b_0+b_1+1}=\bar{1}, n_{b_0+b_1+b_2}=1\};W^{(2)} \rangle $$
$$ \left.  +  \sqrt{1-{b_0\over N}}
|\{n_{b_0+1}=\bar{1},n_{b_0+b_1}=1, n_{b_0+b_1+b_2}=1\};W^{(2)} \rangle \right].$$
Clearly, the terms on the diagonal reproduce the interactions we obtained from a string
attached to a single giant. The off diagonal terms, which correspond to the interactions
in which the giant that the string is attached to is swapped, display a ${1\over r}$ dependence.

\subsection{Boundstate of three AdS giants}

The hop off interaction for a boundstate of three AdS giants is
$$ H|\{ n_1= \overline{a_0+a_1+a_2}, n_2=a_0+a_1, n_3 = a_0 ;W^{(1)} \rangle=$$
$$ 
\lambda\left[ \sqrt{1+{a_0+a_1+a_2\over N}}  A_a
|\{ n_1= \overline{a_0+a_1+a_2+1}, n_2=a_0+a_1, n_3 = a_0 ;W^{(2)} \rangle \right.$$
$$ + \sqrt{1+{a_0+a_1-1\over N}}
B_a |\{ n_1= a_0+a_1+a_2, n_2=\overline{a_0+a_1+1}, n_3 = a_0 ;W^{(2)} \rangle $$
$$ \left.  +  \sqrt{1+{a_0-2\over N}} C_a
|\{ n_1= a_0+a_1+a_2, n_2=a_0+a_1, n_3 = \overline{a_0+1} ;W^{(2)} \rangle \right],$$
$$ $$
$$ $$
$$ H|\{ n_1=a_0+a_1+a_2, n_2=\overline{a_0+a_1}, n_3 = a_0 ;W^{(1)} \rangle=$$
$$ 
\lambda\left[ \sqrt{1+{a_0+a_1+a_2\over N}}  D_a
|\{ n_1= \overline{a_0+a_1+a_2+1}, n_2=a_0+a_1, n_3 = a_0 ;W^{(2)} \rangle \right.$$
$$ + \sqrt{1+{a_0+a_1-1\over N}}
E_a |\{ n_1= a_0+a_1+a_2, n_2=\overline{a_0+a_1+1}, n_3 = a_0 ;W^{(2)} \rangle $$
$$ \left.  +  \sqrt{1+{a_0-2\over N}} F_a
|\{ n_1= a_0+a_1+a_2, n_2=a_0+a_1, n_3 = \overline{a_0+1} ;W^{(2)} \rangle \right],$$
$$ $$
$$ $$
$$ H|\{ n_1= a_0+a_1+a_2, n_2=a_0+a_1, n_3 = \overline{a_0} ;W^{(1)} \rangle=$$
$$ 
\lambda\left[ \sqrt{1+{a_0+a_1+a_2\over N}}  G_a
|\{ n_1= \overline{a_0+a_1+a_2+1}, n_2=a_0+a_1, n_3 = a_0 ;W^{(2)} \rangle \right.$$
$$ + \sqrt{1+{a_0+a_1-1\over N}}
H_a |\{ n_1= a_0+a_1+a_2, n_2=\overline{a_0+a_1+1}, n_3 = a_0 ;W^{(2)} \rangle $$
$$ \left.  +  \sqrt{1+{a_0-2\over N}} I_a
|\{ n_1= a_0+a_1+a_2, n_2=a_0+a_1, n_3 = \overline{a_0+1} ;W^{(2)} \rangle \right],$$
where
$$ A_a=-{\sqrt{a_1+a_2+1}\sqrt{a_1+a_2+3}\sqrt{a_2}\sqrt{a_2+2}\over (a_2+1)(a_1+a_2+2)},$$
$$ B_a=\sqrt{a_1+2\over a_1+1}\sqrt{a_1+a_2+1\over a_1+a_2+2}{1\over a_2+1},$$
$$ C_a=\sqrt{a_2\over a_2+1}\sqrt{a_1\over a_1+1}{1\over a_1+a_2+2},$$
$$ D_a=-\sqrt{a_1\over a_1+1}\sqrt{a_1+a_2+3\over a_1+a_2+2}{1\over a_2+1},$$
$$ E_a=-{\sqrt{a_1}\sqrt{a_1+2}\sqrt{a_2}\sqrt{a_2+2}\over (a_2+1)(a_1+1)},$$
$$ F_a=\sqrt{a_2+2\over a_2+1}\sqrt{a_1+a_2+1\over a_1+a_2+2}{1\over a_1+1},$$
$$ G_a=-\sqrt{a_1+2\over a_1+1}\sqrt{a_2+2\over a_2+1}{1\over a_1+a_2+2},$$
$$ H_a=-\sqrt{a_2\over a_2+1}\sqrt{a_1+a_2+3\over a_1+a_2+2}{1\over a_1+1},$$
$$ I_a-{\sqrt{a_1+a_2+1}\sqrt{a_1+a_2+3}\sqrt{a_1}\sqrt{a_1+2}\over (a_1+1)(a_1+a_2+2)} .$$ 

In the effective field theory limit, the hop off interaction becomes
$$ H|\{ n_1= \overline{a_0+a_1+a_2}, n_2=a_0+a_1, n_3 = a_0 ;W^{(1)} \rangle=$$
$$ 
\lambda\left[ -\sqrt{1+{a_0+a_1+a_2\over N}}
|\{ n_1= \overline{a_0+a_1+a_2+1}, n_2=a_0+a_1, n_3 = a_0 ;W^{(2)} \rangle \right.$$
$$ + \sqrt{1+{a_0+a_1\over N}}{1\over a_2}
|\{ n_1= a_0+a_1+a_2, n_2=\overline{a_0+a_1+1}, n_3 = a_0 ;W^{(2)} \rangle $$
$$ \left.  +  \sqrt{1+{a_0-2\over N}} {1\over a_1+a_2}
|\{ n_1= a_0+a_1+a_2, n_2=a_0+a_1, n_3 = \overline{a_0+1} ;W^{(2)} \rangle \right],$$
$$ $$
$$ $$
$$ H|\{ n_1=a_0+a_1+a_2, n_2=\overline{a_0+a_1}, n_3 = a_0 ;W^{(1)} \rangle=$$
$$ 
\lambda\left[ -\sqrt{1+{a_0+a_1+a_2\over N}} {1\over a_2}
|\{ n_1= \overline{a_0+a_1+a_2+1}, n_2=a_0+a_1, n_3 = a_0 ;W^{(2)} \rangle \right.$$
$$ - \sqrt{1+{a_0+a_1 \over N}}
|\{ n_1= a_0+a_1+a_2, n_2=\overline{a_0+a_1+1}, n_3 = a_0 ;W^{(2)} \rangle $$
$$ \left.  +  \sqrt{1+{a_0\over N}} {1\over a_1}
|\{ n_1= a_0+a_1+a_2, n_2=a_0+a_1, n_3 = \overline{a_0+1} ;W^{(2)} \rangle \right],$$
$$ $$
$$ $$
$$ H|\{ n_1= a_0+a_1+a_2, n_2=a_0+a_1, n_3 = \overline{a_0} ;W^{(1)} \rangle=$$
$$ 
-\lambda\left[ \sqrt{1+{a_0+a_1+a_2\over N}}  {1\over a_1+a_2}
|\{ n_1= \overline{a_0+a_1+a_2+1}, n_2=a_0+a_1, n_3 = a_0 ;W^{(2)} \rangle \right.$$
$$ + \sqrt{1+{a_0+a_1 \over N}}{1\over a_1}
|\{ n_1= a_0+a_1+a_2, n_2=\overline{a_0+a_1+1}, n_3 = a_0 ;W^{(2)} \rangle $$
$$ \left.  +  \sqrt{1+{a_0\over N}}
|\{ n_1= a_0+a_1+a_2, n_2=a_0+a_1, n_3 = \overline{a_0+1} ;W^{(2)} \rangle \right].$$

\subsection{Boundstate of four sphere giants}

The hop off interaction for a boundstate of four sphere giants is
%
%
$$ H|n_{b_0}=1,n_{b_0+b_1}=1,n_{b_0+b_1+b_2}=1,n_{b_0+b_1+b_2+b_3}=\bar{1};W^{(1)}\rangle =$$
$$-\lambda\left[a_s|n_{b_0}=1,n_{b_0+b_1}=1,n_{b_0+b_1+b_2}=1,n_{b_0+b_1+b_2+b_3+1}=\bar{1};W^{(2)}\rangle\right. $$
$$+ b_s |n_{b_0}=1,n_{b_0+b_1}=1,n_{b_0+b_1+b_2+1}=\bar{1},n_{b_0+b_1+b_2+b_3}=1;W^{(2)}\rangle $$
$$+ c_s |n_{b_0}=1,n_{b_0+b_1+1}=\bar{1},n_{b_0+b_1+b_2}=1,n_{b_0+b_1+b_2+b_3}=1;W^{(2)}\rangle $$
$$+\left. d_s |n_{b_0+1}=\bar{1},n_{b_0+b_1}=1,n_{b_0+b_1+b_2}=1,n_{b_0+b_1+b_2+b_3}=1;W^{(2)}\rangle\right], $$
$$ $$
$$ $$
$$ H|n_{b_0}=1,n_{b_0+b_1}=1,n_{b_0+b_1+b_2}=\bar{1},n_{b_0+b_1+b_2+b_3}=1;W^{(1)}\rangle =$$
$$-\lambda\left[e_s|n_{b_0}=1,n_{b_0+b_1}=1,n_{b_0+b_1+b_2}=1,n_{b_0+b_1+b_2+b_3+1}=\bar{1};W^{(2)}\rangle\right. $$
$$+ f_s |n_{b_0}=1,n_{b_0+b_1}=1,n_{b_0+b_1+b_2+1}=\bar{1},n_{b_0+b_1+b_2+b_3}=1;W^{(2)}\rangle $$
$$+ g_s |n_{b_0}=1,n_{b_0+b_1+1}=\bar{1},n_{b_0+b_1+b_2}=1,n_{b_0+b_1+b_2+b_3}=1;W^{(2)}\rangle $$
$$+\left. h_s |n_{b_0+1}=\bar{1},n_{b_0+b_1}=1,n_{b_0+b_1+b_2}=1,n_{b_0+b_1+b_2+b_3}=1;W^{(2)}\rangle\right], $$
$$ $$
$$ $$
$$ H|n_{b_0}=1,n_{b_0+b_1}=\bar{1}, n_{b_0+b_1+b_2}=1, n_{b_0+b_1+b_2+b_3}=1;W^{(1)}\rangle =$$
$$-\lambda\left[i_s|n_{b_0}=1,n_{b_0+b_1}=1,n_{b_0+b_1+b_2}=1,n_{b_0+b_1+b_2+b_3+1}=\bar{1};W^{(2)}\rangle\right. $$
$$+ j_s |n_{b_0}=1,n_{b_0+b_1}=1,n_{b_0+b_1+b_2+1}=\bar{1},n_{b_0+b_1+b_2+b_3}=1;W^{(2)}\rangle $$
$$+ k_s |n_{b_0}=1,n_{b_0+b_1+1}=\bar{1},n_{b_0+b_1+b_2}=1,n_{b_0+b_1+b_2+b_3}=1;W^{(2)}\rangle $$
$$+\left. l_s |n_{b_0+1}=\bar{1},n_{b_0+b_1}=1,n_{b_0+b_1+b_2}=1,n_{b_0+b_1+b_2+b_3}=1;W^{(2)}\rangle\right], $$
$$ $$
$$ $$
$$ H|n_{b_0}=\bar{1},n_{b_0+b_1}=1,n_{b_0+b_1+b_2}=1,n_{b_0+b_1+b_2+b_3}=1;W^{(1)}\rangle =$$
$$-\lambda\left[m_s|n_{b_0}=1,n_{b_0+b_1}=1,n_{b_0+b_1+b_2}=1,n_{b_0+b_1+b_2+b_3+1}=\bar{1};W^{(2)}\rangle\right. $$
$$+ n_s |n_{b_0}=1,n_{b_0+b_1}=1,n_{b_0+b_1+b_2+1}=\bar{1},n_{b_0+b_1+b_2+b_3}=1;W^{(2)}\rangle $$
$$+ o_s |n_{b_0}=1,n_{b_0+b_1+1}=\bar{1},n_{b_0+b_1+b_2}=1,n_{b_0+b_1+b_2+b_3}=1;W^{(2)}\rangle $$
$$+\left. p_s |n_{b_0+1}=\bar{1},n_{b_0+b_1}=1,n_{b_0+b_1+b_2}=1,n_{b_0+b_1+b_2+b_3}=1;W^{(2)}\rangle\right], $$
where the coefficients in the above expressions, together with their effective field theory limit are
$$ a_s=-\sqrt{1-{b_0+b_1+b_2+b_3\over N}}\sqrt{C^1_{b_3}C^1_{b_2+b_3+1}C^1_{b_2+b_3+b_1+2}}\to -\sqrt{1-{b_0+b_1+b_2+b_3\over N}},$$
$$ b_s=\sqrt{1-{b_0+b_1+b_2-1\over N}}\sqrt{b_1+b_2+b_3+2\over b_1+b_2+b_3+3}\sqrt{b_2+b_3+1\over b_2+b_3+2}\sqrt{b_1+b_2+3\over b_1+b_2+2}
\sqrt{b_2+2\over b_2+1}{1\over b_3+1}$$
$$\to \sqrt{1-{b_0+b_1+b_2\over N}}{1\over b_3},$$
$$ c_s=\sqrt{1-{b_0+b_1-2\over N}}\sqrt{b_1+b_2+b_3+2\over b_1+b_2+b_3+3}\sqrt{b_3\over b_3+1}\sqrt{b_1+2\over b_1+1}
\sqrt{b_2\over b_2+1}{1\over b_2+b_3+2}$$
$$\to \sqrt{1-{b_0+b_1\over N}}{1\over b_2+b_3},$$
$$ d_s=\sqrt{1-{b_0-3\over N}}\sqrt{b_2+b_3+1\over b_2+b_3+2}\sqrt{b_1+b_2+1\over b_1+b_2+2}\sqrt{b_1\over b_1+1}
\sqrt{b_3\over b_3 + 1}{1\over b_1+b_2+b_3+3},$$
$$\to \sqrt{1-{b_0\over N}}{1\over b_1+b_2+b_3},$$
$$ e_s=-\sqrt{1-{b_0+b_1+b_2+b_3\over N}}\sqrt{b_1+b_2+b_3+4\over b_1+b_2+b_3+3}\sqrt{b_2\over b_2+1}
\sqrt{b_1+b_2+1\over b_1+b_2+2}\sqrt{b_2+b_3+3\over b_2+b_3+2}{1\over b_3+1}$$
$$\to -\sqrt{1-{b_0+b_1+b_2+b_3\over N}}{1\over b_3},$$
$$ f_s=-\sqrt{1-{b_0+b_1+b_2-1\over N}}\sqrt{C^1_{b_2}C^1_{b_3}C^1_{b_1+b_2+1}}
\to -\sqrt{1-{b_0+b_1+b_2\over N}},$$
$$ g_s=\sqrt{1-{b_0+b_1-2\over N}}\sqrt{b_1+b_2+1\over b_1+b_2+2}\sqrt{b_3+2\over b_3+1}\sqrt{b_1+2\over b_1+1}
\sqrt{b_2+b_3+1\over b_2+b_3+2}{1\over b_2+1}$$
$$\to \sqrt{1-{b_0+b_1\over N}}{1\over b_2},$$
$$ h_s=\sqrt{1-{b_0-3\over N}}\sqrt{b_3+2\over b_3+1}\sqrt{b_2\over b_2+1}\sqrt{b_1\over b_1+1}
\sqrt{b_1+b_2+b_3+2\over b_1+b_2+b_3+3}{1\over b_1+b_2+2}$$
$$\to \sqrt{1-{b_0\over N}}{1\over b_1+b_2},$$
$$ i_s=-\sqrt{1-{b_0+b_1+b_2+b_3\over N}}\sqrt{b_1+b_2+b_3+4\over b_1+b_2+b_3+3}\sqrt{b_1\over b_1+1}
\sqrt{b_2+2\over b_2+1}\sqrt{b_3+2\over b_3+1}{1\over b_2+b_3+2}$$
$$\to -\sqrt{1-{b_0+b_1+b_2+b_3\over N}}{1\over b_2+b_3},$$
$$ j_s=-\sqrt{1-{b_0+b_1+b_2-1\over N}}\sqrt{b_1\over b_1+1}\sqrt{b_2+b_3+3\over b_2+b_3+2}\sqrt{b_3\over b_3+1}
\sqrt{b_1+b_2+2\over b_1+b_2+3}{1\over b_2+1}$$
$$\to -\sqrt{1-{b_0+b_1+b_2\over N}}{1\over b_2},$$
$$ k_s=-\sqrt{1-{b_0+b_1-2\over N}}\sqrt{C^1_{b_1}C^1_{b_2}C^1_{b_2+b_3+1}}\to -\sqrt{1-{b_0+b_1\over N}},$$
$$ l_s=\sqrt{1-{b_0-3\over N}}\sqrt{b_2+2\over b_2+1}\sqrt{b_2+b_3+3\over b_2+b_3+2}\sqrt{b_1+b_2+1\over b_1+b_2+2}
\sqrt{b_1+b_2+b_3+2\over b_1+b_2+b_3+3}{1\over b_1+1},$$
$$\to \sqrt{1-{b_0\over N}}{1\over b_1},$$
$$ m_s=-\sqrt{1-{b_0+b_1+b_2+b_3\over N}}\sqrt{b_1+2\over b_1+1}\sqrt{b_1+b_2+3\over b_1+b_2+2}
\sqrt{b_3+2\over b_3+1}\sqrt{b_2+b_3+3\over b_2+b_3+2}{1\over b_1+b_2+b_3+3}$$
$$\to -\sqrt{1-{b_0+b_1+b_2+b_3\over N}}{1\over b_1+b_2+b_3},$$
$$ n_s=-\sqrt{1-{b_0+b_1+b_2-1\over N}}\sqrt{b_1+2\over b_1+1}\sqrt{b_1+b_2+b_3+4\over b_1+b_2+b_3+3}\sqrt{b_3\over b_3+1}
\sqrt{b_2+2\over b_2+1}{1\over b_1+b_2+2}$$
$$\to -\sqrt{1-{b_0+b_1+b_2\over N}}{1\over b_1+b_2},$$
$$ o_s=-\sqrt{1-{b_0+b_1-2\over N}}\sqrt{b_1+b_2+3\over b_1+b_2+2}\sqrt{b_1+b_2+b_3+4\over b_1+b_2+b_3+3}\sqrt{b_2\over b_2+1}
\sqrt{b_2+b_3+1\over b_2+b_3+2}{1\over b_1+1}$$
$$\to -\sqrt{1-{b_0+b_1\over N}}{1\over b_1},$$
$$ p_s=-\sqrt{1-{b_0-3\over N}}\sqrt{C^1_{b_1}C^1_{b_1+b_2+1}C^1_{b_1+b_2+b_3+2}}\to -\sqrt{1-{b_0\over N}}.$$

\subsection{Boundstate of four AdS giants}

The hop off interaction for a boundstate of four sphere giants is
%
%
$$ H|n_1=\overline{a_0+a_1+a_2+a_3}, n_2=a_0+a_1+a_2, n_3=a_0+a_1, n_4=a_0; W^{(1)}\rangle =$$
$$-\lambda\left[a_a|n_1=\overline{a_0+a_1+a_2+a_3+1}, n_2=a_0+a_1+a_2, n_3=a_0+a_1, n_4=a_0;W^{(2)}\rangle\right. $$
$$+ b_a |n_1=a_0+a_1+a_2+a_3, n_2=\overline{a_0+a_1+a_2+1}, n_3=a_0+a_1, n_4=a_0;W^{(2)}\rangle $$
$$+ c_a | n_1=a_0+a_1+a_2+a_3, n_2=a_0+a_1+a_2, n_3=\overline{a_0+a_1+1}, n_4=a_0;W^{(2)}\rangle $$
$$+\left. d_a | n_1=a_0+a_1+a_2+a_3, n_2=a_0+a_1+a_2, n_3=a_0+a_1, n_4=\overline{a_0+1};W^{(2)}\rangle\right], $$
$$ $$
$$ $$
$$ H|n_1=a_0+a_1+a_2+a_3, n_2=\overline{a_0+a_1+a_2}, n_3=a_0+a_1, n_4=a_0;W^{(1)}\rangle =$$
$$-\lambda\left[e_a|n_1=\overline{a_0+a_1+a_2+a_3+1}, n_2=a_0+a_1+a_2, n_3=a_0+a_1, n_4=a_0;W^{(2)}\rangle\right. $$
$$+ f_a |n_1=a_0+a_1+a_2+a_3, n_2=\overline{a_0+a_1+a_2+1}, n_3=a_0+a_1, n_4=a_0;W^{(2)}\rangle $$
$$+ g_a |n_1=a_0+a_1+a_2+a_3, n_2=a_0+a_1+a_2, n_3=\overline{a_0+a_1+1}, n_4=a_0;W^{(2)}\rangle $$
$$+\left. h_a | n_1=a_0+a_1+a_2+a_3, n_2=a_0+a_1+a_2, n_3=a_0+a_1, n_4=\overline{a_0+1};W^{(2)}\rangle\right], $$
$$ $$
$$ $$
$$ H|n_1=a_0+a_1+a_2+a_3, n_2=a_0+a_1+a_2, n_3=\overline{a_0+a_1}, n_4=a_0;W^{(1)}\rangle =$$
$$-\lambda\left[i_a|n_1=\overline{a_0+a_1+a_2+a_3+1}, n_2=a_0+a_1+a_2, n_3=a_0+a_1, n_4=a_0;W^{(2)}\rangle\right. $$
$$+ j_a |n_1=a_0+a_1+a_2+a_3, n_2=\overline{a_0+a_1+a_2+1}, n_3=a_0+a_1, n_4=a_0;W^{(2)}\rangle $$
$$+ k_a |n_1=a_0+a_1+a_2+a_3, n_2=a_0+a_1+a_2, n_3=\overline{a_0+a_1+1}, n_4=a_0;W^{(2)}\rangle $$
$$+\left. l_a | n_1=a_0+a_1+a_2+a_3, n_2=a_0+a_1+a_2, n_3=a_0+a_1, n_4=\overline{a_0+1};W^{(2)}\rangle\right], $$
$$ $$
$$ $$
$$ H|n_1=a_0+a_1+a_2+a_3, n_2=a_0+a_1+a_2, n_3=a_0+a_1, n_4=\overline{a_0}; W^{(1)}\rangle =$$
$$-\lambda\left[m_a|n_1=\overline{a_0+a_1+a_2+a_3+1}, n_2=a_0+a_1+a_2, n_3=a_0+a_1, n_4=a_0;W^{(2)}\rangle\right. $$
$$+ n_a |n_1=a_0+a_1+a_2+a_3, n_2=\overline{a_0+a_1+a_2+1}, n_3=a_0+a_1, n_4=a_0;W^{(2)}\rangle $$
$$+ o_a |n_1=a_0+a_1+a_2+a_3, n_2=a_0+a_1+a_2, n_3=\overline{a_0+a_1+1}, n_4=a_0;W^{(2)}\rangle $$
$$+\left. p_a | n_1=a_0+a_1+a_2+a_3, n_2=a_0+a_1+a_2, n_3=a_0+a_1, n_4=\overline{a_0+1};W^{(2)}\rangle\right], $$
where the coefficients in the above expressions, together with their effective field theory limit are
$$ a_a=\sqrt{1+{a_0+a_1+a_2+a_3\over N}}\sqrt{C^1_{a_3}C^1_{a_2+a_3+1}C^1_{a_2+a_3+a_1+2}}\to \sqrt{1+{a_0+a_1+a_2+a_3\over N}},$$
$$ b_a=-\sqrt{1+{a_0+a_1+a_2-1\over N}}\sqrt{a_1+a_2+a_3+2\over a_1+a_2+a_3+3}\sqrt{a_2+a_3+1\over a_2+a_3+2}\sqrt{a_1+a_2+3\over a_1+a_2+2}
\sqrt{a_2+2\over a_2+1}{1\over a_3+1}$$
$$\to -\sqrt{1+{a_0+a_1+a_2\over N}}{1\over a_3},$$
$$ c_a=-\sqrt{1+{a_0+a_1-2\over N}}\sqrt{a_1+a_2+a_3+2\over a_1+a_2+a_3+3}\sqrt{a_3\over a_3+1}\sqrt{a_1+2\over a_1+1}
\sqrt{a_2\over a_2+1}{1\over a_2+a_3+2}$$
$$\to -\sqrt{1+{a_0+a_1\over N}}{1\over a_2 + a_3},$$
$$ d_a=-\sqrt{1+{a_0-3\over N}}\sqrt{a_2+a_3+1\over a_2+a_3+2}\sqrt{a_1+a_2+1\over a_1+a_2+2}\sqrt{a_1\over a_1+1}
\sqrt{a_3\over a_3+1}{1\over a_1+a_2+a_3+3},$$
$$\to -\sqrt{1+{a_0\over N}}{1\over a_1+a_2+a_3},$$
$$ e_a=\sqrt{1+{a_0+a_1+a_2+a_3\over N}}\sqrt{a_1+a_2+a_3+4\over a_1+a_2+a_3+3}\sqrt{a_2\over a_2+1}
\sqrt{a_1+a_2+1\over a_1+a_2+2}\sqrt{a_2+a_3+3\over a_2+a_3+2}{1\over a_3+1}$$
$$\to \sqrt{1+{a_0+a_1+a_2+a_3\over N}}{1\over a_3},$$
$$ f_a=\sqrt{1+{a_0+a_1+a_2-1\over N}}\sqrt{C^1_{a_2}C^1_{a_3}C^1_{a_1+a_2+1}}
\to \sqrt{1+{a_0+a_1+a_2\over N}},$$
$$ g_a=-\sqrt{1+{a_0+a_1-2\over N}}\sqrt{a_1+a_2+1\over a_1+a_2+2}\sqrt{a_3+2\over a_3+1}\sqrt{a_1+2\over a_1+1}
\sqrt{a_2+a_3+1\over a_2+a_3+2}{1\over a_2+1}$$
$$\to -\sqrt{1+{a_0+a_1\over N}}{1\over a_2},$$
$$ h_a=-\sqrt{1+{a_0-3\over N}}\sqrt{a_3+2\over a_3+1}\sqrt{a_2\over a_2+1}\sqrt{a_1\over a_1+1}
\sqrt{a_1+a_2+a_3+2\over a_1+a_2+a_3+3}{1\over a_1+a_2+2}$$
$$\to -\sqrt{1+{a_0\over N}}{1\over a_1+a_2},$$
$$ i_a=\sqrt{1+{a_0+a_1+a_2+a_3\over N}}\sqrt{a_1+a_2+a_3+4\over a_1+a_2+a_3+3}\sqrt{a_1\over a_1+1}
\sqrt{a_2+2\over a_2+1}\sqrt{a_3+2\over a_3+1}{1\over a_2+a_3+2}$$
$$\to \sqrt{1+{a_0+a_1+a_2+a_3\over N}}{1\over a_2+a_3},$$
$$ j_a=\sqrt{1+{a_0+a_1+a_2-1\over N}}\sqrt{a_1\over a_1+1}\sqrt{a_2+a_3+3\over a_2+a_3+2}\sqrt{a_3\over a_3+1}
\sqrt{a_1+a_2+2\over a_1+a_2+3}{1\over a_2+1}$$
$$\to \sqrt{1+{a_0+a_1+a_2\over N}}{1\over a_2},$$
$$ k_a=\sqrt{1+{a_0+a_1-2\over N}}\sqrt{C^1_{a_1}C^1_{a_2}C^1_{a_2+a_3+1}}\to \sqrt{1+{a_0+a_1\over N}},$$
$$ l_a=-\sqrt{1+{a_0-3\over N}}\sqrt{a_2+2\over a_2+1}\sqrt{a_2+a_3+3\over a_2+a_3+2}\sqrt{a_1+a_2+1\over a_1+a_2+2}
\sqrt{a_1+a_2+a_3+2\over a_1+a_2+a_3+3}{1\over a_1+1},$$
$$\to -\sqrt{1+{a_0\over N}}{1\over a_1},$$
$$ m_a=\sqrt{1+{a_0+a_1+a_2+a_3\over N}}\sqrt{a_1+2\over a_1+1}\sqrt{a_1+a_2+3\over a_1+a_2+2}
\sqrt{a_3+2\over a_3+1}\sqrt{a_2+a_3+3\over a_2+a_3+2}{1\over a_1+a_2+a_3+3}$$
$$\to \sqrt{1+{a_0+a_1+a_2+a_3\over N}}{1\over a_1+a_2+a_3},$$
$$ n_a=\sqrt{1+{a_0+a_1+a_2-1\over N}}\sqrt{a_1+2\over a_1+1}\sqrt{a_1+a_2+a_3+4\over a_1+a_2+a_3+3}\sqrt{a_3\over a_3+1}
\sqrt{a_2+2\over a_2+1}{1\over a_1+a_2+2}$$
$$\to \sqrt{1+{a_0+a_1+a_2\over N}}{1\over a_1+a_2},$$
$$ o_a=\sqrt{1+{a_0+a_1-2\over N}}\sqrt{a_1+a_2+3\over a_1+a_2+2}\sqrt{a_1+a_2+a_3+4\over a_1+a_2+a_3+3}\sqrt{a_2\over a_2+1}
\sqrt{a_2+a_3+1\over a_2+a_3+2}{1\over a_1+1}$$
$$\to \sqrt{1+{a_0+a_1\over N}}{1\over a_1},$$
$$ p_a=\sqrt{1+{a_0-3\over N}}\sqrt{C^1_{a_1}C^1_{a_1+a_2+1}C^1_{a_1+a_2+a_3+2}}\to \sqrt{1+{a_0\over N}}.$$

\section{Dual Description for AdS Giants}

In section 3.1.2 we have associated a lattice state to each Young diagram. 
Recall that we denote the number of boxes in row $i$ by $r_i$. 
The lattice state corresponding to a particular Young diagram is obtained by interpreting $n_i=r_i-r_{i+1}\ge 0$
as an occupation number for lattice site $i$. Note that we set $r_{N+1}=0$. 
This description is simple because

\begin{itemize}

\item{} The lattice is not dynamical; it has a fixed total of $N$ sites. This is a consequence of the fact that a Young
        diagram has at most $N$ rows.

\item{} At leading order at large $N$, the number of particles on the lattice when describing a bound state of sphere giants is fixed.

\end{itemize} 

This description is less convenient for the description of a bound state of AdS giants, because even at leading order 
in the large $N$ limit, the
number of particles on the lattice fluctuates. In this appendix we will suggest an alternative lattice description of the AdS giant
bound states. We have not yet developed this description in detail.
The advantage of the new description, is that at leading order, in the large $N$ limit, we obtain a description for which
the lattice is not dynamical and the number of particles on the lattice is again fixed.

Let $c_i$ denote the number of boxes in the $i$th column.
Our alternative lattice description is obtained by setting the occupation number of lattice site $i$ equal to
$$ n_i=c_i-c_{i+1}. $$
There are two features of this new lattice description that we would like to stress

\begin{itemize}

\item{} Since there is no bound on the number of columns in a Young diagram, it is best to phrase the description in terms of the lattice 
        obtained by accounting only for the occupied columns. The number of occupied columns can change, so that this leads to a dynamical
        lattice description. 

\item{} One can have at most $N$ boxes in any column, implying that there is a bound of $N$ on the number of particles that can occupy any given
        lattice site. 

\end{itemize}

The total $U(1)$
${\cal R}$ charge $K$ which is equal to the number of $Z$s in the giant plus the number of $Z$s in the string, is fixed. 
By placing all of the $Z$s in the first row we can reach at most the $K$th lattice site. Thus, we can
work on a lattice of $\le K$ sites. Further, at leading order in the large $N$
limit, processes that change the number of rows in the Young diagram are suppressed, so that the number of particles occupying
the lattice is fixed at leading order. 
The bound on the number of particles allowed to occupy any given site is easily accounted for by employing the $q$-deformed 
algebra\cite{qalg}
$$ \hat{a}^\dagger |n\rangle = \sqrt{\big[ n+1\big]}|n+1\rangle , \qquad \hat{a}|n\rangle=\sqrt{\big[ n\big]}|n-1\rangle ,$$
with
$$ \big[ n\big]\equiv {1-q^n\over 1-q},\qquad q =e^{2\pi i\over N+1}.$$
We need to use this representation for both the Cuntz oscillators $\hat{A}_i$ and the open string operators $\hat{W}_i$, since the bound
is on the sum of particle plus open string number. For $O(1)$ AdS giants (which is the case we have in mind), this bound on the number
of particles in each site can be neglected and we can use the usual Cuntz oscillators.

The lattice state is described by listing the occupation numbers for the occupied sites. The site occupied by the open string is indicated
with a bar. Thus, as an example of the new notation, we have
$$\young({\,}{\,}{\,}{\,}{\,}{\,}{\,}{\,}{w},{\,}{\,}{\,}{\,})\leftrightarrow \{ n_4=1, n_9=\bar{1}\} .$$
Using this new lattice description, the hop off interaction for a boundstate of two AdS giants, for example, can be expressed as
$$ H|\{ n_{a_0+a_1}=\overline{1},n_{a_0}=1\};W^{(1)}\rangle =
-\lambda\left[\sqrt{1+{a_0+a_1\over N}}
\sqrt{C_{a_1}^1}|\{ n_{a_0+a_1+1}=\overline{1},n_{a_0}=1 \};W^{(2)}\rangle\right.$$
$$+\left. \sqrt{1+{a_0\over N}}{1\over a_1+1}|\{ n_{a_0+a_1}=1,n_{a_0+1}=\overline{1}\};W^{(2)}\rangle\right],$$
$$ H|\{n_{a_0+a_1}=1, n_{a_0}=\overline{1} \};W^{(1)} \rangle =-\lambda\left[\sqrt{1+{a_0\over N}}
\sqrt{C^1_{a_1}}|\{n_{a_0+a_1}=1, n_{a_0+1}=\overline{1}\}; W^{(2)} \rangle\right.$$
$$-\left. \sqrt{1+{a_0+a_1\over N}}{1\over a_1+1}|\{n_{a_0+a_1+1}=
\overline{1},n_{a_0}=1\};W^{(2)} \rangle\right].
$$

\end{document}